\documentclass[12pt]{article}
\pdfoutput=1
\usepackage[utf8]{inputenc}
\usepackage[DIV13]{typearea}
\usepackage{ragged2e}
\usepackage{amsmath,amsfonts,bbm,mathrsfs,amssymb}
\usepackage{slashed,cancel}
\usepackage[usenames,dvipsnames]{xcolor}
\usepackage{cite}
\usepackage{bm} 
\usepackage{hyperref}
\hypersetup{colorlinks=true,urlcolor=Magenta,anchorcolor=blue,citecolor=blue,filecolor=blue,
            linkcolor=Magenta,menucolor=blue, linktocpage=true,pdfproducer=medialab}
\usepackage[english]{babel}
\usepackage{catchfile}
\usepackage{indentfirst}
\usepackage{graphicx}
\usepackage{float}
\usepackage{enumerate}
\usepackage{subcaption}
\usepackage{multirow}
\usepackage{tikz-feynman}
\usepackage[normalem]{ulem} 

\newcommand{\email}[1]{\href{mailto:#1}{\tt #1}}

\numberwithin{equation}{section}

\newcommand{\blue}[1]{\color{blue} #1 \color{black}}

\newcommand{\magenta}[1]{\color{Magenta} #1 \color{black}}
\newcommand{\be}{\begin{equation}}
\newcommand{\ee}{\end{equation}}
\newcommand{\ba} {\begin{equation}\begin{aligned}}
\newcommand{\ea} {\end{aligned}\end{equation}}
\newcommand{\bea}{\begin{eqnarray}}
\newcommand{\eea}{\end{eqnarray}}

\newcommand{\LL}{\mathscr{L}}

\newcommand{\OO}{\mathcal{O}}

\newcommand{\absval}[1]{\left| #1 \right|}

\newcommand{\hc}{\text{h.c.}}
\newcommand{\ov}[1]{\overline{#1}}
\newcommand{\nn}{\nonumber}

\newcommand{\TeV}{\ \text{TeV}}
\newcommand{\GeV}{\ \text{GeV}}
\newcommand{\MeV}{\ \text{MeV}}

\def\diag{{\tt diag}}

\def\sm{{\rm SM}}
\def\BR{{\rm BR}}

\def\Re{{\texttt{Re}}}
\def\Im{{\texttt{Im}}}

\newcommand{\p}[1]{\left(#1\right)}

\begin{document}
\renewcommand*{\thefootnote}{\fnsymbol{footnote}}
\begin{titlepage}

\vspace*{-1cm}
\flushleft{\magenta{FTUAM-21-3}}
\hfill{\magenta{IFT-UAM/CSIC-21-94}}
\\[1cm]

\begin{center}
\bf\LARGE \blue{
Searching for BSM Physics in Yukawa Couplings and Flavour Symmetries}\\[4mm]
\centering
\vskip .3cm
\end{center}
\vskip 0.5  cm
\begin{center}
{\large\bf J.~Alonso-Gonz\'alez}${}^{a)}$~\footnote{\email{j.alonso.gonzalez@csic.es}},
{\large\bf A.~de Giorgi}${}^{a)}$~\footnote{\email{arturo.degiorgi@uam.es}},\\[2mm]
{\large\bf L.~Merlo}${}^{a)}$~\footnote{\email{luca.merlo@uam.es}}, and 
{\large\bf S.~Pokorski}${}^{b)}$~\footnote{\email{Stefan.Pokorski@fuw.edu.pl}}
\vskip .7cm
{\footnotesize
${}^{a)}$ Departamento de F\'isica Te\'orica and Instituto de F\'isica Te\'orica UAM/CSIC,\\
Universidad Aut\'onoma de Madrid, Cantoblanco, 28049, Madrid, Spain
\vskip .2cm
${}^{b)}$ Institute of Theoretical Physics, Faculty of Physics,\\ 
University of Warsaw, Pasteura 5, PL 02-093, Warsaw, Poland
}
\end{center}
\vskip 2cm
\begin{abstract}
\justify 

In the framework of the Standard Model Effective Field Theory, we compare  the lower bounds on the scale of new physics possibly contributing to the $f\bar{f}h$ effective couplings, obtained from the measurements of different observables, under the assumption that the Wilson coefficients of the relevant dim 6 operators respect certain flavour structure:  either the Minimal Flavour Violation (MFV) ansatz or a flavour symmetry, often invoked  to explain  the observed pattern of fermion masses and mixings. We perform a global analysis of the bounds following from  the limits on the diagonal couplings measured in the Higgs boson production and decays  at the LHC experiments. Another set of  bounds is obtained from the  limits on non-diagonal couplings constrained by the variety of  flavour changing neutral current (FCNC)  and radiative decay processes. With the present precision of the LHC data, the FCNC data give stronger bounds on the scale of new physics than the collider data (obviously, for the MFV ansatz only collider data are relevant): once the Wilson coefficients respect some flavour structure, the obtained bounds are in the TeV range. In the quark case, these limits are compatible with a few percent deviations from the SM Yukawa couplings and only mildly more stringent than those obtained  from the available collider data. For leptons, instead, the FCNC bounds are stronger and then a signal in the near future collider data would mean the violation of the flavour symmetry or indicate the presence of additional beyond the Standard Model contributions, affecting the flavour observables, that leads to cancellations.

\end{abstract}
\end{titlepage}
\setcounter{footnote}{0}

\pdfbookmark[1]{Table of Contents}{tableofcontents}
\tableofcontents

\renewcommand*{\thefootnote}{\arabic{footnote}}

\section{Introduction}
\label{sec:intro}

So far, no signal of beyond the Standard Model (SM) particles has been found in direct searches. Therefore, indirect searches of new physics (NP) effects in high  precision experiments may play similarly important role as the precision electroweak (EW) measurements played in constraining the top quark and Higgs boson masses. 
A useful tool for interpreting the precision measurements is the effective field theory (EFT) approach, where higher dimension operators are added to the SM Lagrangian. In this  paper we investigate in the framework of the   Standard Model Effective Field Theory (SMEFT)~\cite{Buchmuller:1985jz,Grzadkowski:2010es} approach~\footnote{ An alternative approach that will not be considered here is the so-called Higgs Effective Field Theory~\cite{Feruglio:1992wf,Grinstein:2007iv,Contino:2010mh,Alonso:2012px,Alonso:2012pz,Buchalla:2013rka,Brivio:2013pma,Brivio:2014pfa,Gavela:2014vra,Alonso:2014wta,Hierro:2015nna,Gavela:2016bzc,Brivio:2016fzo,Merlo:2016prs} that  treats the physical Higgs field and the SM Goldstone bosons as independent degrees of freedom.} the discovery potential  for NP effects in  the Higgs boson Yukawa couplings  in different observables under the assumption that the Wilson coefficients of the relevant dim 6 operators respect certain flavour structure. 

There are several experimental sources of information about potential contributions of beyond the SM (BSM) physics to the  Higgs boson couplings to fermions. One is the Higgs boson production and decays at colliders. So far, only flavour conserving decays have been observed. An increase in the precision of such measurements is one of the main goals of the future collider experimental programmes. One may even hope that in the future Higgs factories  theoretically interesting bounds on flavour violating decays  can be obtained, or they can be discovered. Another source are the low energy observables: magnetic (MDM) and electric dipole moments (EDM) (sensitive to the diagonal Yukawa couplings) and a large variety of flavour changing neutral current (FCNC) processes, including radiative  decays and the CP symmetry violating ones. If BSM physics gives rise to flavour violating Higgs couplings, those processes are sensitive to them.

In the SMEFT approach, at the level of dimension six operators, there is only one operator contributing to the Yukawa couplings of the structure ${(\ov{f}}Hf)HH^\dagger$  , with three $3\times3$ matrices in  the flavour space of the Wilson coefficients.  Each measured observable puts bounds on some combination of the Wilson coefficients expressed in the fermion mass eigenstate basis. 

It is a common procedure to use the bounds on the Wilson coefficients of higher dimension operators added to the SM Lagrangian for estimating sensitivity of various observables to NP scales. For instance, very recently, the European Particle Physics Strategy committee has provided a qualitative illustration of the sensitivity to the NP scales in the present and future experiments based on the  contributions of  dim 6 operators, one at a time and with $\OO(1)$ couplings in the Wilson coefficients, to various observables, see Fig.~5.1 in Ref.~\cite{EuropeanStrategyforParticlePhysicsPreparatoryGroup:2019qin}. It has been shown in Ref.~\cite{EuropeanStrategyforParticlePhysicsPreparatoryGroup:2019qin} that,  putting aside  several  (important) caveats of that analysis, in principle the flavour  precision-low energy observables can be  sensitive to much higher NP scales than the  collider observables. 

We investigate and compare the sensitivity to NP scale in both mentioned above sets of experimental data, that is collider and flavour data, under the assumption that the Wilson coefficients are  either  consistent with the Minimal Flavour Violation (MFV) hypothesis~\cite{Chivukula:1987py,DAmbrosio:2002vsn,Cirigliano:2005ck,Davidson:2006bd,Alonso:2011jd} or respect  a flavour symmetry, often invoked to explain the structure of fermion masses and mixing. In UV complete models extending  the SM, once the flavour symmetry is imposed, it should also manifest itself in the structure of the Wilson coefficients of the operators contributing to the Yukawa couplings in the SMEFT framework. This fact has a strong  impact on the lower bounds on the NP scales derived from the experimental data since the couplings in the Wilson coefficients are controlled by the symmetry. With the Wilson coefficients respecting a flavour symmetry,  we investigate the complementarity of different processes in establishing the most stringent bounds, that is  of their sensitivity to NP scales. The emerging picture is different from the one obtained with uncontrolled Wilson coefficients,  providing valuable indications on where to look for NP that is related to the Higgs sector. 

In the first part of the paper we systematically reanalyse the bounds on the diagonal Higgs couplings following from the electron EDM (nuclear, atomic and molecular EDMs are not taken into account due to the large uncertainties affecting their theoretical predictions and the ambiguities associated to their analyses~\cite{Chien:2015xha,Cirigliano:2016nyn}) and from the Higgs boson production and decays at the LHC. In the second part we use the bounds on the off-diagonal couplings following from the FCNC processes, to investigate their complementarity with the collider bounds, once  a flavour symmetry is assumed. In conclusions we summarise  the emerging  picture for the lower bounds on the NP scales,  much more optimistic from the point of view of direct searches for new  particles than what follows  when no flavour symmetry is invoked.

Within the same framework, it has recently been shown~\cite{Alonso-Gonzalez:2021jsa} that the bounds from the electron EDM on CP violation in the electron Yukawa coupling can be translated into bounds on CP violation in tau Yukawa coupling that are two orders of magnitude stronger than if no flavour symmetry is considered, with strong implications for the electroweak baryogenesis. On the other side, the idea of linking the diagonal to off-diagonal Higgs-quark couplings with the purpose  of constraining the diagonal light generation Yukawas by the tree-level contribution of the off-diagonal ones to the $K$, $B$ and $D$ meson oscillations has been proposed in Ref.~\cite{Goertz:2014qia}. In that paper, the misalignment between the quark mass and the Yukawa matrices is introduced by  adding $d=6$ operators with anarchical flavour structure to the SM Yukawa coupling. The diagonal and off-diagonal couplings are related by arbitrary coefficients that are randomly generated within a specific interval in a numerical analysis. Moreover, in the recent paper Ref.~\cite{Calibbi:2021qto}, the implications of the muon $g-2$ results on the lepton mass matrix have been studied considering effective $d=6$ dipole operators. See also Refs.~\cite{He:2015rqa,Baek:2016pef}, where a bunch of $d=6$ SMEFT operators have been considered in the MFV context to discuss Lepton Flavour Violating Higgs decays.

The structure of the paper can be seen in the table of Contents.

\section{The Effective Lagrangian}
\label{sec:EffLag}

Generic NP that affects the Yukawa couplings can be encoded at low energies into effective operators. The relevant part of the Lagrangian can be written within the SMEFT approach as
\begin{equation}
\begin{split}
\LL= &-\ov{Q_L'}\tilde H Y'_u u'_R-\ov{Q_L'} H Y'_d d'_R -\ov{L_L'} H Y'_e e'_R + \\
&-\left(\ov{Q_L'}\tilde H C'_u u'_R + \ov{Q_L'} H C'_d d'_R\right)\dfrac{H^\dag H}{\Lambda_q^2} 
- \ov{L_L'} H C'_e e'_R\dfrac{H^\dag H}{\Lambda_\ell^2}+\hc\,,
\end{split}
\label{Non-RenormLag}
\end{equation}
where the prime denotes fields and quantities in the flavour basis. $Q_L'$ and $L_L'$ are respectively the quark and lepton EW doublets, $u'_R$, $d'_R$ and $e'_R$ stand for the EW singlets and $H$ is the Higgs EW doublet. The scales $\Lambda_{q,\ell}$ are the effective scales of NP, in general different for the quark and lepton sectors. Finally $Y'_f$ and $C'_f$ are $3\times 3$ complex matrices in the flavour space. Notice that flavour indices are omitted.

After the electroweak symmetry breaking (EWSB), the Lagrangian in Eq.~\eqref{Non-RenormLag} can be rewritten as
\begin{equation}
\begin{split}
\LL= &-\left[\ov{u'_L}\left(Y'_u+\dfrac{v^2}{2\Lambda_q^2}C'_u\right) u'_R 
+ \ov{d'_L} \left(Y'_d+\dfrac{v^2}{2\Lambda_q^2}C'_d\right) d'_R 
+ \ov{e'_L} \left(Y'_e+\dfrac{v^2}{2\Lambda_\ell^2}C_e'\right) e'_R\right]\dfrac{v}{\sqrt2} + \\
&-\left[\ov{u'_L}\left(Y'_u+\dfrac{3v^2}{2\Lambda_q^2}C'_u\right) u'_R
+ \ov{d'_L} \left(Y'_d+\dfrac{3v^2}{2\Lambda_q^2}C'_d\right) d'_R
+ \ov{e'_L} \left(Y'_e+\dfrac{3v^2}{2\Lambda_\ell^2}C'_e\right) e'_R\right]\dfrac{h}{\sqrt2} + \\
& +\hc+\ldots\,,
\end{split}
\label{LagH}
\end{equation}
where $v=246\GeV$ is the EW symmetry breaking vacuum expectation value (VEV) of the Higgs field. The first line gives the fermion mass terms, and the second line describes the fermion interactions with the physical Higgs field $h$. Dots stand for terms with more than one Higgs field that will not be discussed in this paper.

The  effective Yukawa matrices are diagonalised by bi-unitary rotations involving the left- and right-handed fields:
\begin{equation}
Y'_f+\dfrac{v^2}{2\Lambda^2}C'_f=V_f Y_f U_f^\dag \, ,
\label{Rotations}
\end{equation}
where $V_f$ and $U_f$ are unitary $3\times3$ matrices. Here and in the following, $f=u,d,e$ refer to the up and down quark sectors, and to the lepton sector, respectively.  The matrices $V_f$ enter into the definitions of the CKM and PMNS matrices. In this expression and the following ones where no index is explicitly indicated, $\Lambda$ should be identified with $\Lambda_q$ when $f=u,d$, and with $\Lambda_\ell$ when $f=e$. The matrices $Y_f$ are diagonal matrices whose entries are defined in terms of fermion masses:
\be
\begin{gathered}
Y_u= \frac{\sqrt2}{v}M_u= \frac{\sqrt2}{v} \left(m_u,m_c, m_t\right) \,, \\
Y_d= \frac{\sqrt2}{v}M_d= \frac{\sqrt2}{v} \left(m_d,m_s, m_b\right) \, , \\
Y_e= \frac{\sqrt2}{v}M_e= \frac{\sqrt2}{v}  \left(m_e,m_\mu, m_\tau\right) \, ,
\end{gathered}
\label{diagonalYuk}
\end{equation}
whose experimental determination is~\cite{Zyla:2020zbs}
\begin{align}
m_u(2\GeV)&=2.16^{+0.49}_{-0.26}\MeV
\quad
&&m_c(m_c)=1.27\pm0.02\GeV
\quad
&& \begin{cases}
m_t(m_t)=162.5^{+2.1}_{-1.5}\GeV\\
m_t^\text{pole}=172.76\pm0.30\GeV\\
\end{cases}\nn\\
m_d(2\GeV)&=4.67^{+0.48}_{-0.17}\MeV
\quad
&&m_s(2\GeV)=93^{+11}_{-5}\MeV
\quad
&&m_b(m_b)=4.18^{+0.03}_{-0.02}\GeV\nn\\
m_e&=0.511\MeV
\quad
&&m_\mu=105.66\MeV
\quad
&&m_\tau=1776.86\pm0.12\MeV\,,
\end{align}
where the relative error on the electron (muon) mass is of $1.96\times 10^{-10}$ ($2.27\times 10^{-8}$). Notice that for the $u$, $d$, and $s$ quark mass values refer to $m_q(\mu=2\GeV)$ in the $\ov{\text{MS}}$ mass-independent subtraction scheme; the $c$ and $b$ quark mass values are the running masses in the $\ov{\text{MS}}$ scheme, $m_q(m_q)$; the two values listed for the $t$ quark mass refer to the running mass in the $\ov{\text{MS}}$ scheme and the pole mass. In order to determine the value of the Yukawa couplings, the $\ov{\text{MS}}$ mass value for the top has been adopted.

Finally, the Lagrangian in the mass eigenstate basis  reads
\begin{equation}
\begin{split}
\LL=
& -\ov{u_L}Y_u u_R\dfrac{v}{\sqrt2}
-\ov{d_L} Y_d d_R \dfrac{v}{\sqrt2}
-\ov{e_L} Y_e e_R\dfrac{v}{\sqrt2} + \\
& -\left[\ov{u_L}\left(Y_u+\dfrac{v^2}{\Lambda_q^2}C_u\right) u_R 
+ \ov{d_L} \left(Y_d+\dfrac{v^2}{\Lambda_q^2}C_d\right) d_R
+ \ov{e_L} \left(Y_e +\dfrac{v^2}{\Lambda_\ell^2}C_e\right) e_R\right]\dfrac{h}{\sqrt2}+ \\
& +\hc+\ldots\,,
\end{split}
\label{LagHMassBasis}
\end{equation}
where
\begin{equation}
C_f= V_f^\dag C'_f U_f
\label{CC' matrix}
\end{equation}
are the Wilson coefficients of the fermion-Higgs operators in this basis. It will be useful in the following to introduce the following notation for the effective fermion-Higgs couplings:
\be
\hat Y_f \equiv Y_f+\dfrac{v^2}{\Lambda^2}C_f\,.
\label{hatYf}
\ee

The effective Lagrangian below the EWSB scale, in the mass eigenstate basis, describing flavour conserving Higgs-fermion couplings, that is usually introduced in the literature is~\cite{Peskin:2013xra,Lepage:2014fla,Brod:2013cka}
\be
\LL_\text{eff}=-\dfrac{y_\psi}{\sqrt2}\left(\kappa_\psi \ov{\psi}\psi+i\tilde{\kappa}_{\psi}\ov\psi\gamma_5\psi\right)h\,,
\label{Brod_Lag}
\ee
where $\psi$ refers to a specific quark or charged lepton, $y_\psi=\sqrt2 m_\psi/v$, $m_\psi$ is the $\psi$ mass. The $\kappa_\psi$ and $\tilde \kappa_\psi$ are real numbers parametrising the CP conserving and CP violating parts of these couplings. In the SM limit, $\kappa_\psi=1$ and $\tilde \kappa_\psi=0$. Matching this effective Lagrangian with the one in Eq.~\eqref{LagHMassBasis}, we  conclude that
\begin{equation}
Y_f K_f=Y_f + \dfrac{v^2}{\Lambda^2}\diag(\texttt{Re} C_f) \, , \qquad
Y_f {\tilde K_f}=\dfrac{v^2}{\Lambda^2}\diag(\texttt{Im} C_f)\,,
\label{Kappa_matrices}
\end{equation}
where $K_f$ is the diagonal matrix containing the $\kappa_\psi$ parameters. For instance, for the charged leptons: \mbox{$ K_e\equiv\diag(\kappa_e,\,\kappa_\mu,\,\kappa_\tau)$}, and similarly for the quarks. When, in Sect.~\ref{sec:OffDiag}, we will discuss the phenomenology associated to the off-diagonal entries of $\hat Y_f$ in Eq.~\eqref{hatYf}, we will abandon the $\kappa$-notation and explicitly use the entries $\hat{y}_{\psi\psi'}$ of the effective Yukawa $\hat Y_f$, as customary done in the literature.

We also introduce a parameter that will be useful in discussing the constraints on the diagonal couplings of the Higgs with fermions: this parameter measures the deviation from the SM predicted value of the Yukawa couplings and can be defined as
\be
r^2_\psi\equiv\dfrac{v^2|\hat{y}_{\psi\psi}|^2}{2 m^2_\psi}=\kappa_\psi^2+\tilde\kappa_\psi^2\,.
\label{rpsiDef}
\ee
On the right-hand side of the previous expression, we express the parameter $r_\psi$ in terms of the $\kappa_\psi$ and $\tilde{\kappa}_\psi$.
\\

Once a flavour symmetry is in action to describe fermion masses and mixings, fermions transform under such a symmetry determining the flavour structure of both $Y'_f$ and $C'_f$. Although these two matrices may be independent of  each other, they own in general the same flavour structure: it is this fact that allows to link the different entries of $\hat Y_f $. We will consider in the next subsection two benchmark scenarios that are representative of what occurs in general when flavour symmetries act at the Lagrangian level. 

\subsection{Flavour Structures}
\label{sec:FlavourContexts}

\subsubsection*{Minimal Flavour Violation}
The first framework we will consider is the so-called Minimal Flavour Violation (MFV)~\cite{Chivukula:1987py,DAmbrosio:2002vsn,Cirigliano:2005ck,Davidson:2006bd,Alonso:2011jd}: according to it, any source of flavour and CP violation within the SM and beyond is encoded into the Yukawa couplings of the SM. The modern description of the MFV consists in considering fermions transforming non-trivially under a product of $U(3)$ symmetries, one for each fermion Weyl spinors. The Yukawa terms are made invariant under this symmetry by promoting the Yukawa couplings to spurion fields transforming as bi-triplets of the flavour symmetry. Fermion masses and mixings are then described (but not explained as pointed out in Refs.~\cite{Albrecht:2010xh,Alonso:2011yg,Alonso:2012fy,Alonso:2013mca}) once these spurions acquire a proper background value.
 
The strength of the MFV resides in lowering the scale of NP that can give rise to generic higher order operators describing flavour observables. Indeed, any higher dimensional operator encoding flavour changing effects needs insertions of the Yukawa spurions in order to be invariant under the flavour symmetry. Once the spurions acquire their background values, the Wilson coefficients of these operators are suppressed by factors proportional to fermion masses and mixings. It follows that present experimental bounds from different experiments, both low-energy flavour factories and high-energy particle colliders, allow NP contributions with a cut-off of a few TeV~\cite{DAmbrosio:2002vsn,Cirigliano:2005ck,Davidson:2006bd,Alonso:2011jd,Grinstein:2010ve,Feldmann:2010yp,Guadagnoli:2011id,Buras:2011zb,Redi:2011zi,Buras:2011wi,Alonso:2012jc,Lopez-Honorez:2013wla,Alonso:2016onw,Dinh:2017smk,Arias-Aragon:2017eww,Merlo:2018rin,Arias-Aragon:2020qip,Arias-Aragon:2022ats}, instead of hundred of TeV if no flavour structure is imposed in the Wilson coefficients~\cite{Isidori:2010kg}.

Here we apply the MFV ansatz to the dim 6 operators contributing to the  Yukawa couplings. Implementing the MFV construction in the Lagrangian in Eq.~\eqref{Non-RenormLag} leads to
\be
C'_f=c'_fY'_f
\ee
where $c'_f$ are flavour blind complex coefficients with moduli of $\OO(1)$. It follows that fermion Yukawas and fermion-Higgs couplings are aligned: both are diagonal  in the  fermion mass eigenstate basis, with the eigenvalues of the latter rescaled by the factor $(1+v^2c^\prime/\Lambda^2)$. Therefore, in the MFV framework, the Higgs boson does not acquire flavour violating couplings to fermions and, moreover, the $\kappa_\psi$ and $\tilde{\kappa}_\psi$ are flavour independent in each sector:
\begin{equation}
\begin{gathered}
\kappa_t=\kappa_c=\kappa_u \simeq 1+\dfrac{v^2}{\Lambda_q^2}\texttt{Re}\,c'_u\, , \qquad\qquad
\tilde\kappa_t=\tilde\kappa_c=\tilde\kappa_u \simeq \dfrac{v^2}{\Lambda_q^2}\texttt{Im}\,c'_u\, ,
\end{gathered}
\label{k_psi_relations_MFV}
\end{equation}
and similarly for the down-quarks and leptons.
A useful relation is the one connecting $\kappa^2_\psi$ and $\tilde{\kappa}^2_\psi$,
\be
(\kappa_\psi-1)^2+\tilde{\kappa}_\psi^2 \simeq |c^\prime_f|^2\frac{v^4}{\Lambda^4},
\label{kappas_relation}
\ee
for any fermion $\psi$ in the sector $f$.
Finally, the explicit expression for $r_\psi$ turns out to be
\be
r^2_t=r^2_c=r^2_u \simeq 1+\dfrac{v^4}{\Lambda_q^4}|c'_u|^2+2\dfrac{v^2}{\Lambda_q^2}\texttt{Re}\,c'_u\,,
\label{ruMFV}
\ee 
and similarly for the down-quarks and the charged leptons.

\subsubsection*{Froggatt-Nielsen}
The second benchmark framework we consider is the Froggatt-Nielsen (FN) one~\cite{Froggatt:1978nt}, that is the simplest flavour model one can think of in terms of spectrum and dimension of the flavour symmetry. Indeed, the latter is a global Abelian $U(1)$ factor and the spectrum is the SM one augmented by an additional scalar field $\phi$, singlet under the SM gauge symmetries and transforming non-trivially only under the $U(1)$. Fermions may transform under this flavour symmetry and the Yukawa terms are made invariant under it by the insertion of powers of the scalar field, originating non-renormalisable operators.  

Taking for concreteness, but without any loss of generality, the $U(1)$-charges of the left-handed (LH) adjoint fields and of the right-handed (RH) fields as positive integers and the $U(1)$-charge for the scalar $\phi$ as $n_\phi=-1$, the Yukawa Lagrangian then reads
\begin{equation}
\begin{split}
\LL_\text{FN}= &
-y'_{u,ij}\,\ov{Q'_{Li}} \tilde H u'_{Rj}\left(\dfrac{\phi}{\Lambda_F}\right)^{(n_{Q_i}+n_{u_j})}
-y'_{d,ij}\,\ov{Q'_{Li}} H d'_{Rj}\left(\dfrac{\phi}{\Lambda_F}\right)^{(n_{Q_i}+n_{d_j})} + \\
& -y'_{e,ij}\,\ov{L'_{Li}} H e'_{Rj}\left(\dfrac{\phi}{\Lambda_F}\right)^{(n_{L_i}+n_{e_j})} + \\ 
& -\Bigg[c'_{u,ij}\,\ov{L'_{Li}} \tilde H u'_{Rj} \left(\dfrac{\phi}{\Lambda_F}\right)^{(n_{Q_i}+n_{u_j})}
+ c'_{d,ij}\,\ov{L'_{Li}} H d'_{Rj}\left(\dfrac{\phi}{\Lambda_F}\right)^{(n_{Q_i}+n_{d_j})}\Bigg]\dfrac{H^\dag H}{\Lambda_q^2} + \\ 
& -c'_{e,ij}\,\ov{L'_{Li}} H e'_{Rj}\left(\dfrac{\phi}{\Lambda_F}\right)^{(n_{L_i}+n_{e_j})}\dfrac{H^\dag H}{\Lambda_\ell^2} + \hc \,,
\end{split}
\label{Non-RenormLagFN}
\end{equation}
where $i$ and $j$ are flavour indices, and $y'_{f,ij}$ and $c'_{f,ij}$ are free complex parameter with moduli of order $\OO(1)$. The cut-off scale $\Lambda_F$ is different from the scales $\Lambda_{q,\ell}$ that suppresses the $d=6$ operators in Eq.~\eqref{Non-RenormLag}: these scales indeed correspond to two different sectors of the theory and may have nothing in common. 

Once the scalar field $\phi$ acquires VEV, the different operators get suppressed by powers of the ratio
\be
\epsilon \equiv \langle\phi\rangle/\Lambda_F\,.
\ee 
Typically $n_{Q_1}>n_{Q_2}>n_{Q_3}$, $n_{u_1}>n_{u_2}>n_{u_3}$ and similarly for the remaining fermion Weyl spinors~\footnote{For a discussion of various constraints on the charge assignment in the FN models see, for instance, Refs.~\cite{Dudas:1995yu,Chankowski:2005qp,Altarelli:2002sg,Altarelli:2012ia,Bergstrom:2014owa} and references therein.}: this guarantees the correct charged fermion mass hierarchies and small mixings. While the latter is the general expectation for the CKM matrix, it is not a very strong requirement in the lepton sector, but follows from the prejudice that the large mixings of the PMNS matrix arise mainly from the neutrino sector. The condition imposed on the charges guarantees a simple parametrisation of the mixing angles. According to the definition in Eq.~\eqref{Rotations}, for the up sector we get:
\begin{gather}
Y_u=\diag(y_u\,\epsilon^{n_{Q_1}+n_{u_1}},\,y_c\,\epsilon^{n_{Q_2}+n_{u_2}},\,y_t \,\epsilon^{n_{Q_3}+n_{u_3}}) \nn \\
V_{u,ij}= \delta_{ij} + (1-\delta_{ij})\frac{(j-i)}{|j-i|}\frac{y_{u,ij}}{y_{u, jj}}\epsilon^{|{n_{Q_i}-n_{Q_j}}|} \\
U_{u,ij}= \delta_{ij} + (1-\delta_{ij})\frac{(j-i)}{|j-i|}\frac{y_{u, ji}}{y_{u, jj}}\epsilon^{|{n_{u_i}-n_{u_j}}|} \,,\nn
\label{diagonalFNandRots}
\end{gather}
and similarly for the down-quark and charged lepton sectors with the corresponding substitutions. $y_{u,ij}$ and $y_i$ are complex and real parameters of $\OO(1)$, respectively.

In the mass eigenstate basis, the Higgs boson does have flavour changing couplings with fermions and the diagonal ones have the same suppressions as the Yukawa couplings. The matrices $C_f$  given by Eq.~\eqref{CC' matrix}, using the expressions for $C'_f$, $V_f$ and $U_f$ above, read:
\begin{gather}
C_{u,ij}\approx \OO(1)\epsilon^{n_{Q_i}+n_{u_j}}e^{i\theta_{u,ij}}\,,\quad
C_{d,ij}\approx \OO(1)\epsilon^{n_{Q_i}+n_{d_j}}e^{i\theta_{d,ij}}\,,\quad
C_{e,ij}\approx \OO(1)\epsilon^{n_{L_i}+n_{e_j}}e^{i\theta_{e,ij}}\,,\nn\\
C_{f,ii}\approx \OO(Y_{f,i})e^{i\theta_{f,ii}}\,,
\label{ExplicitCFN}
\end{gather}
where $\theta_{f,ij}$ are effective phases resulting from the product in Eq.~\eqref{CC' matrix}. Once focussing on the diagonal elements, the resulting expressions for the $\kappa_\psi$ and $\tilde{\kappa}_\psi$ then read
\be
\begin{aligned}
K_f=&1 + \dfrac{v^2}{\Lambda^2}\diag\left(\OO(1)\cos \theta_{f,11},\,\OO(1)\cos \theta_{f,22},\,\OO(1)\cos \theta_{f,33}\right) \, ,\\
\tilde K_f=&\dfrac{v^2}{\Lambda^2}\diag\left(\OO(1)\sin \theta_{f,11},\,\OO(1)\sin \theta_{f,22},\,\OO(1)\sin \theta_{f,33}\right) \,,
\end{aligned}
\label{KfKf}
\ee
where $\OO(1)$ factors reflect the presence of the free coefficients associated to each effective operator and typical of the FN construction. Notice the difference with the MFV scenario, where the angle $\theta$ is flavour independent.
Instead, similarly to Eq.~\eqref{kappas_relation} in the MFV case, we have
\be
(\kappa_\psi-1)^2+\tilde{\kappa}_\psi^2=\mathcal{O}(1)^2\frac{v^4}{\Lambda^4}\,.
\label{kappas_relation1}
\ee
where  $\OO(1)$'s are due to free flavour dependent parameters. Finally, for $r_\psi$ the  Eq.~\eqref{ruMFV} gets replaced by
\be
r_\psi^2\simeq1+\OO(1)^2\dfrac{v^4}{\Lambda^4}+2\OO(1)\dfrac{v^2}{\Lambda^2}\cos\theta_{f,\psi}\,,
\label{ruFN}
\ee
for each fermion $\psi$.

To fully determine the size of the Yukawa couplings it is necessary to consider a concrete set of FN charges and a value for the parameter $\epsilon$. Tab.~\ref{tab:FNscen} contains possible settings that agree with data once fixing $\epsilon=0.23$. For the lepton sector, three distinct models have been considered, Anarchy ($A$), $\mu-\tau$-Anarchy ($A_{\mu\tau}$) and Hierarchy ($H$) as studied in Refs.~\cite{Altarelli:2002sg,Altarelli:2012ia,Bergstrom:2014owa}: the first realisation corresponds to the anarchical ansatz for the neutrino Yukawa matrix~\cite{Hall:1999sn,Haba:2000be,deGouvea:2003xe}; the second introduces some degree of hierarchy in the entries associated to the first interacting neutrino; while the last model envisage a hierarchical structure within all the three interacting neutrinos. With the present sensitivity on neutrino oscillation data, the three models are compatible with the measurements, although the hierarchical models have a higher evidence with respect to the anarchical one in a bayesian analysis~\cite{Bergstrom:2014owa}. Interestingly, in the numerical analysis that follows, only the LH field charges are relevant: it is indeed possible to express the dependence on the RH field charges in terms of fermion masses and the LH fermion charges. 

\begin{table}[h!]
\centering
\begin{tabular}{|c|c|c|c|}
\hline
&&&\\[-4mm]
\multirow{2.5}{*}{\bf Quarks}				& $Q'_L$		& $u'_R$ 			& $d'_R$ \\[1mm]
\cline{2-4}&&&\\[-4mm]
								& $(2,1,0)$ 	& $(5,2,0)$		& $(5,4,2)$\\[1mm]
\hline
\end{tabular}\\
\begin{tabular}{|c|c|c|}
\hline
&&\\[-4mm]
{\bf Leptons}						& $L'_L$		& $e'_R$ 					\\[1mm]
\hline
&&\\[-4mm]
Anarchy ($A$) 						& $(0,0,0)$	& $(10,5,3)$		 \\
$\mu\tau$-Anarchy ($A_{\mu\tau}$) 		& $(1,0,0)$	& $(9,5,3)$		 \\
Hierarchy ($H$) 					& $(2,1,0)$	& $(8,4,3)$		 \\[1mm]
\hline
\end{tabular}
\caption{\em FN charge assignments for realistic scenarios with $\epsilon=0.23$.}
\label{tab:FNscen}
\end{table}

\section{Bounds from EDMs and Collider Higgs Boson Data}
\label{sec:Diag}

This section is dedicated to the discussion of the constraints on the diagonal fermion-Higgs couplings. There are two type of constraints that can be taken into consideration: bounds from EDMs and bounds from colliders.

\subsection{Bounds from EDMs}
With some BSM CP violation effects encoded in non-vanishing parameters $\tilde \kappa_\psi$, one obtains contributions to the electron EDM from the Barr-Zee diagrams with the fermions $\psi$ running in the loop: neglecting the $Z$-mediated contribution, the expression for the eEDM is given by~\cite{Brod:2013cka}
\be
\frac{d_e}{e}=4 \, N_C \, Q^2_\psi \, \dfrac{\alpha_\text{em}}{(4\pi)^3} \, \sqrt2  \, G_F \, m_e\left[\kappa_e\tilde\kappa_\psi f_1(x_{\psi/h})+\tilde\kappa_e\kappa_\psi f_2(x_{\psi/h})\right]
\label{EqeEDM}
\ee
where $Q_\psi$ is the $\psi$ electric charge, $N_C=3$ if the fermion $\psi$ running in the loop is a quark and $N_C=1$ if it is a charged lepton, $\alpha_\text{em}$ is the fine structure constant at the scale of the electron mass, $G_F$ the Fermi constant, and $f_1$ and $f_2$ are functions of $x_{\psi/h}=m_\psi^2/m_h^2$, defined by
\be
\begin{aligned}
f_1(x)&=\dfrac{2x}{\sqrt{1-4x}}\left[Li_2\Bigg(1-\dfrac{1-\sqrt{1-4x}}{2x}\right)-Li_2\left(1-\dfrac{1+\sqrt{1-4x}}{2x}\right)\Bigg]\\
f_2(x)&=(1-2x)f_1(x)+2x(\ln x+2)\,,
\end{aligned}
\ee
where $Li_2$  is the usual dilogarithm
\be
Li_2(x)=-\int_0^x du \dfrac{\ln(1-u)}{u}\,.
\ee

Recent studies showed that the present experimental bound on the electric EDM from the ACME II collaboration~\cite{Andreev:2018ayy}
\begin{equation}
|d_e|<1.1\times 10^{-29}\text{ e cm} \, , \qquad  \text{at }90\%\text{ C.L.} \,,
\end{equation}
infer upper limits on the $\tilde\kappa_\psi$ parameters for several fermions: assuming only one third generation fermion running in the loop at a time, we find at $90\%\text{ C.L.}$
\be
|\tilde\kappa_t|\lesssim 0.0012\,,\qquad
|\tilde\kappa_b|\lesssim 0.24\,,\qquad
|\tilde\kappa_\tau|\lesssim 0.29\,,
\label{Boundkappatildetau1}
\ee
that agree with the results presented in Ref.~\cite{Fuchs:2020uoc}~\footnote{Ref.~\cite{Fuchs:2020uoc} based their calculus on Ref.~\cite{Panico:2018hal}, where the formula in Eq.~(2.33) should be corrected including a factor $1/\sqrt{2}$ according to Ref.~\cite{Brod:2013cka}.}. Notice that the top quark mass used is the running one. The bound we found for $|\tilde\kappa_b|$ is smaller than the value found in Ref.~\cite{Brod:2018pli}, $|\tilde\kappa_b|\lesssim 0.32$: the reason of the difference is the level of the confidence level of the result. 

Moreover, considering a quark charm and a muon in the loop with modified couplings with the Higgs, we get the following results at $90\%\text{ C.L.}$
\be
|\tilde\kappa_c|\lesssim 0.37\,, \qquad
|\tilde\kappa_\mu|\lesssim 31\,.
\label{BoundkappatildeCharmMuon}
\ee
The values obtained in Ref.~\cite{Brod:2018pli} for the charm quark, $|\tilde\kappa_c|<0.82$, is larger than what we find and the reason is the different confidence level at which the two bounds are given. 

Finally, assuming that only the $e\ov eh$ Yukawa coupling received contributions from NP, the Barr-Zee contribution to the electron EDM gives very strong bound on $\tilde\kappa_e$~\cite{Alonso-Gonzalez:2021jsa} that at $90\%\text{ C.L.}$ reads
\be
|\tilde\kappa_e|\lesssim 0.0017\,.
\label{Boundkappatildee}
\ee

As reported in Ref.~\cite{Alonso-Gonzalez:2021jsa}, once the presence of an underlying flavour symmetry is assumed, the bound in Eq.~\eqref{Boundkappatildee} can be translated into one on $|\tilde\kappa_\tau|$: at $90\%\text{ C.L.}$
\be
|\tilde\kappa_\tau|\lesssim \alpha\,0.0017\,,
\label{Boundkappatildetau2}
\ee
with $\alpha=1$ in the MFV case and $\alpha=\OO(1)$ in the FN one. Although this result has been obtained assuming NP contributions only to the charged lepton sector, it approximatively holds also when such contributions to all the Yukawa couplings are allowed (under the hypothesis of the absence of 1:1000 cancellation among the electron and top contributions). This result is two order of magnitudes stronger than the one listed in Eq.~\eqref{Boundkappatildetau1}. 

Before moving on to the new section and analysing bounds from colliders, let us discuss the impact of this result on the original parameter space. When $\tilde\kappa_\psi$ is strongly constrained, there are two possible explanations: either the corresponding phase $\theta_{f, ii}$ is extremely small by accident or due to a symmetry, or the cut-off scale $\Lambda$ is large. In this second case, 
\be
\begin{cases}
|\tilde\kappa_t|\lesssim 0.0012&\Longrightarrow \Lambda_q\gtrsim7.4\TeV\\[2mm]
|\tilde\kappa_e|,|\tilde\kappa_\tau|\lesssim 0.0017&\Longrightarrow \Lambda_\ell\gtrsim6.0\TeV\,,
\end{cases}
\label{BoundsLambdaEDMs}
\ee
taking the $\OO(1)$ coefficients  equal to $1$.

\subsection{Bounds From Colliders}
The LHC signal strength of the Higgs production by a mechanism $P$ with the Higgs decaying into a final state $F$ , normalised to the SM prediction, is defined as
\be
\mu_P^F = 
\frac{\sigma_P}{\sigma^\text{SM}_P}\, 
\frac{\Gamma\p{h \rightarrow F}}{\Gamma^\text{SM}\p{h \rightarrow F}}\, 
\p{\frac{\Gamma_{h,\text{tot}}}{\Gamma^\text{SM}_{h,\text{tot}}}}^{-1} \,,
\ee
where $\sigma_P$ is the production cross section by the mechanism  $P$, $\Gamma\p{h \rightarrow F}$ is the decay rate into the final state $F$ and $\Gamma_{h,\text{tot}}$ is the total decay rate of the Higgs. The ``SM'' prefix indicates that the quantity is the predicted one within the SM.  The parameter controlling the deviation from the SM expectation in the Higgs signal strength is $r_\psi^2$ introduced in Eq.~\eqref{rpsiDef}.

The most recent combinations of different Higgs signal strengths has been released by both ATLAS and CMS collaborations in Refs.~\cite{ATLAS:2019nkf} and \cite{CMS:2018uag}, respectively, using data at $\sqrt s=13\TeV$. The results that will be used in this analysis are reported in Tabs.~\ref{tab:ATLASData} and \ref{tab:CMSData}.

\begin{table}[h!]
\centering 
\begin{tabular}{|c|c|c|}
\hline
\multicolumn{3}{|c|}{}\\[-2mm]
\multicolumn{3}{|c|}{\textbf{ATLAS}\qquad $\sqrt s=13\TeV$ \qquad $24.5-79.8\text{ fb}^{-1}$} \\[2mm]
\hline \hline
&&\\[-4mm]
Production mech. $P$  & Final state $F$ & $\sigma\times BR$ normalised to SM \\ \hline
&&\\[-4mm]
\multirow{4}{*}{ggF} 			& $\gamma\gamma$ 		& $0.96\pm0.14$\\
						& $ZZ^*$					& $1.04^{+0.16}_{-0.15}$\\
						& $WW^*$				& $1.08\pm0.19$\\
						& $\tau\ov\tau$				& $0.96^{+0.59}_{-0.52}$\\
\hline
&&\\[-4mm]
\multirow{5}{*}{VBF} 			& $\gamma\gamma$ 		& $1.39^{+0.40}_{-0.35}$\\
						& $ZZ^*$					& $2.68^{+0.98}_{-0.83}$\\
						& $WW^*$				& $0.59^{+0.36}_{-0.35}$\\
						& $\tau\ov\tau$				& $1.16^{+0.58}_{-0.53}$\\
						& $b\ov b$				& $3.01^{+1.67}_{-1.61}$\\
\hline
&&\\[-4mm]
\multirow{3}{*}{VH} 			& $\gamma\gamma$ 		& $1.09^{+0.58}_{-0.54}$\\
						& $ZZ^*$					& $0.68^{+1.20}_{-0.78}$\\
						& $b\ov b$				& $1.19^{+0.27}_{-0.25}$\\
\hline
&&\\[-4mm]
\multirow{4}{*}{$t\ov tH+tH$} 	& $\gamma\gamma$ 		& $1.10^{+0.41}_{-0.35}$\\
						& $VV^*$					& $1.50^{+0.59}_{-0.57}$\\
						& $\tau\ov\tau$				& $1.38^{+1.13}_{-0.96}$\\
						& $b\ov b$				& $0.79^{+0.60}_{-0.59}$\\ 
\hline
\end{tabular}
\caption{\em Higgs signal strength from ATLAS collaboration in Ref.~\cite{ATLAS:2019nkf}.  ggF stands for gluon fusion, VBF for vector boson fusion, VH for vector boson-Higgs boson associated production, $t\ov tH+tH$ for associated Higgs  boson production with $t\ov t$ or single $t$.} 
\label{tab:ATLASData}
\end{table}

\begin{table}[h!]
\centering 
\begin{tabular}{|c|c|c|}
\hline
\multicolumn{3}{|c|}{}\\[-2mm]
\multicolumn{3}{|c|}{\textbf{CMS}\qquad $\sqrt s=13\TeV$ \qquad $35.9\text{ fb}^{-1}$} \\[2mm]
\hline \hline
&&\\[-4mm]
Production mech. $P$  & Final state $F$ & $\sigma\times BR$ normalised to SM \\ \hline
&&\\[-4mm]
\multirow{6}{*}{ggF} 	&$bb$ & $2.51^{+2.43}_{-2.01}$\\
& $\tau\tau$ &$1.05^{+0.53}_{-0.47}$\\
& $WW$ &$1.35^{+0.21}_{-0.19}$\\
& $ZZ$ &$1.22^{+0.23}_{-0.21}$\\
& $\gamma\gamma$ &$1.16^{+0.21}_{-0.18}$\\
& $\mu\mu$ &$0.31^{+1.80}_{-1.79}$\\
\hline
&&\\[-4mm]
\multirow{5}{*}{VBF} 	& $\tau\tau$ &$1.12^{+0.45}_{-0.43}$\\
& $WW$ &$0.28^{+0.64}_{-0.60}$\\
& $ZZ$ &$-0.09^{+1.02}_{-0.76}$\\
& $\gamma\gamma$ &$0.67^{+0.59}_{-0.46}$\\
& $\mu\mu$ &$2.72^{+7.12}_{-7.03}$\\
\hline
&&\\[-4mm]
\multirow{4}{*}{WH} 	&$bb$ &$1.73^{+0.70}_{-0.68}$\\
& $WW$ &$3.91^{+2.26}_{-2.01}$\\
& $ZZ$ &$0.00^{+2.33}_{-0.00}$\\
& $\gamma\gamma$ &$3.76^{+1.48}_{-1.35}$\\
\hline
&&\\[-4mm]
\multirow{4}{*}{ZH} 	&$bb$ &$0.99^{+0.47}_{-0.45}$\\
& $WW$ &$0.96^{+1.81}_{-1.46}$\\
& $ZZ$ &$0.00^{+4.26}_{-0.00}$\\
& $\gamma\gamma$ &$0.00^{+1.14}_{-0.00}$\\
\hline
&&\\[-4mm]
\multirow{5}{*}{$t\ov{t}H$} 	&$bb$ &$0.91^{+0.45}_{-0.43}$\\
& $\tau\tau$ &$0.23^{+1.03}_{-0.88}$\\
& $WW$ &$1.60^{+0.65}_{-0.59}$\\
& $ZZ$ &$0.00^{+1.50}_{-0.00}$\\
& $\gamma\gamma$ &$2.18^{+0.88}_{-0.75}$\\
\hline
\end{tabular}
\caption{\em
Higgs signal strength from CMS collaboration in Ref.~\cite{CMS:2018uag}.  ggF stands for gluon fusion, VBF for vector boson fusion, ZH and WH for Z- and W-Higgs boson associated production, respectively, while  $t\ov tH$ for associated Higgs  boson production with $t\ov t$ pair.}
\label{tab:CMSData}
\end{table}
 
In addition, both collaborations reported in Refs.~\cite{ATLAS:2020fzp} and \cite{CMS:2020xwi} the observations of $h\to\mu\ov\mu$ in collisions at $\sqrt{s}=13\TeV$. The data in Tab.~\ref{tab:ATLASDatamumu} that will be used in our analysis are the ones with associated jet emission and VBF (more details follow), while all the data in Tab.~\ref{tab:CMSDatamumu} will be used.

\begin{table}[h!]
\centering 
\begin{tabular}{|c|c|c|}
\hline
\multicolumn{3}{|c|}{}\\[-2mm]
\multicolumn{3}{|c|}{\textbf{ATLAS}\qquad $h\to\mu\ov\mu$\qquad$\sqrt s=13\TeV$\qquad $139\text{ fb}^{-1}$}
\\[2mm] \hline \hline
&&\\[-4mm]
Production mech. $P$  & Category & $\sigma\times \BR$ normalised to SM \\ \hline
&&\\[-4mm]
\multirow{3}{*}{ggF} 			& $0$-jet 				& $-0.4\pm1.6$\\
						& $1$-jet				& $2.4\pm1.2$\\
						& $2$-jet				& $-0.6\pm1.2$\\
\hline
&&\\[-4mm]
\multirow{1}{*}{VBF} 			& 	 				& $1.8\pm1.0$\\
\hline
&&\\[-4mm]
\multirow{1}{*}{VH+$t\ov{t}H$} 	&		 	 		& $5.0\pm3.5$\\
\hline
\end{tabular}
\caption{\em The Higgs signal strength parameter for $h\to\mu\ov\mu$ from ATLAS collaboration in Ref.~\cite{ATLAS:2020fzp}. The production mechanisms are the ggF, the VBF, and the combination of VH and $t\ov{t}H$; in the first case, the categories considered are with $0$, $1$ and $2$ gluon associated emission.} 
\label{tab:ATLASDatamumu}
\end{table}

\begin{table}[h!]
\centering 
\begin{tabular}{|c|c|}
\hline
\multicolumn{2}{|c|}{}\\[-2mm]
\multicolumn{2}{|c|}{\textbf{CMS}\qquad $h\to\mu\ov\mu$\qquad$\sqrt s=13\TeV$\qquad $137\text{ fb}^{-1}$}
\\[2mm] \hline \hline
&\\[-4mm]
Production mech. $P$   & $\sigma\times \BR$ normalised to SM \\ \hline
&\\[-4mm]
\multirow{1}{*}{ggH+$t\ov{t}H$} 			 	 				& $0.66^{+0.67}_{-0.66}$\\
\hline
&\\[-4mm]
\multirow{1}{*}{VBF+VH} 			 	 		& $1.84^{+0.89}_{-0.77}$\\
\hline
\end{tabular}
\caption{\em The Higgs signal strength parameter for $h\to\mu\ov\mu$ from CMS collaboration in Ref.~\cite{CMS:2020xwi}. The production mechanisms are a combination of ggF and $t\ov{t}H$ in the first row, and of VBF and VH in the second.}
\label{tab:CMSDatamumu}
\end{table}

In Ref.~\cite{CMS:2019hve}, the CMS collaboration reported the most stringent bound on $h\to c\ov{c}$, analysing data on  vector boson-Higgs (VH) associated production  with the Higgs decaying into two charm quarks. The best fit value, $\mu_{cc}=37\pm20$, is however affected by very large uncertainties, much larger than the ones on the previous reported data, and for this reason will not be used in our analysis. 

Let us report the last determination of the total decay width of the Higgs boson presented by the CMS collaboration in Ref.~\cite{CMS:2019ekd}:
\be
\Gamma^\text{CMS}_{h, \text{tot}}=3.2^{+2.8}_{-2.2}\MeV\,,
\label{HiggsTotalWidthExp}
\ee
while the SM prediction is $\Gamma^\text{SM}_{h, \text{tot}}=4.1\MeV$~\cite{LHCHiggsCrossSectionWorkingGroup:2016ypw}.

We can now proceed with the determination of the constraints on the parameters of our effective Lagrangian considering the above data. First we will discuss the case in which only one Yukawa coupling deviates from the corresponding SM prediction at a time, leaving for the end of the section the discussion of NP effects in all the Yukawas couplings simultaneously, and the impact of the action of a flavour symmetry.

\subsubsection{NP In Only One Yukawa Coupling At A Time}

Let us first consider the case in which NP affects only one Yukawa coupling of the fermion $\psi$ at a time, focusing on any fermion but the top quark, which will be discussed separately later on. In this case, the Higgs boson production  cross section by any mechanism  is approximatively the one predicted in the SM :
\be
\dfrac{\sigma_P}{\sigma^\text{SM}_P}=1\,,\qquad \text{for any $P$}\,.
\ee
On the other hand, the partial decay width does deviate from the SM prediction, according to
\be
\dfrac{\Gamma\p{h \rightarrow F}}{\Gamma^\text{SM}\p{h \rightarrow F}}=
\begin{cases}
1\,,\quad&\text{for }F\neq\psi\ov\psi\\
r_\psi^2\,,\quad&\text{for }F=\psi\ov\psi\,.
\end{cases}
\ee
Notice that NP loop contributions are neglected if the SM contributions arise at tree-level, while they are present if the SM contributions arise at their same loop-level.
Finally, the total decay width of the Higgs  boson is given by
\be
\dfrac{\Gamma_{h,\text{tot}}}{\Gamma^\text{SM}_{h,\text{tot}}}=1+\BR^\text{SM}_{\psi\psi}\left(r_\psi^2-1\right)\,,
\label{RatioGammaGammaSMtot}
\ee
where $\BR^\text{SM}_{\psi\psi}$ stand for the Higgs branching ratios into $\psi\ov\psi$ in the SM, whose numerical values are reported in Tab.~\ref{tab:BRSM}.

\begin{table}[h!]
\centering 
{\renewcommand{\arraystretch}{1.5}
\begin{tabular}{|c|c|c|c|c|c|}
\hline
$\BR^\sm_{bb}$ & $\text{BR}^\sm_{gg}$ &$\text{BR}^\sm_{\tau\tau}$ & $\text{BR}^\sm_{cc}$ & $\text{BR}^\sm_{\gamma\gamma}$ &  $\text{BR}^\sm_{\mu\mu}$ \\
\hline
\hline
$0.58$ & $0.082$ & $0.062$ & $0.029$ & $0.002$ & $0.0002$ \\
\hline
\end{tabular}
\caption{\em SM  Branching Ratios for the Higgs  boson decaying into $\psi\ov\psi$, taking $m_h = 125.10\GeV$~\cite{LHCHiggsCrossSectionWorkingGroup:2016ypw}.}
\label{tab:BRSM}}
\end{table}

All in all, the Higgs signal strength assuming that the top-Higgs coupling is strictly the SM one, while allowing NP contributions to other fermion-Higgs couplings, is given by the following expression:
\begin{equation}
\mu_P^F =
\begin{cases}
\dfrac{1}{1 + \text{BR}^\sm_{\psi \psi}\p{r_\psi^2 - 1}} \, , &\qquad \text{for } F \neq \psi\ov\psi \\
\dfrac{r_\psi^2}{1 + \text{BR}^\sm_{\psi \psi}\p{r_\psi^2 - 1}} \, , &\qquad \text{for } F = \psi\ov\psi \, \, ,
\end{cases}
\label{eq:muNoTop}
\end{equation}

The case in which NP enters the top Yukawa coupling differs from the previous one because  the top-loops provide the largest contributions to some of the production mechanisms. Indeed, the explicit expression of the normalised production cross section is given by
\be
\dfrac{\sigma_P}{\sigma^\text{SM}_P} = 
\begin{cases}
1 \, , \qquad &\text{for } P = \text{VBF}, \text{VH} \\
r_t^2 \, , \qquad &\text{for } P = \text{ggF}, \text{ttH}+\text{tH} \,.
\end{cases}
\ee

The partial decay width also deviates from the SM prediction, in particular the channel with $\gamma\gamma$ in the final state is affected:
\be
\frac{\Gamma\p{h \rightarrow F}}{\Gamma^\text{SM}\p{h \rightarrow F}} =
\begin{cases}
1 \, , \qquad &\text{for } F \neq \gamma\gamma \\
1.639-0.718\,r_t
\, , \qquad &\text{for } F = \gamma\gamma \,,
\end{cases}
\ee
where the second expression has been obtained from Ref.~\cite{Choi:2021nql}, accounting for the top- and $W$-loop contributions, and indicating only the dominant linear dependence on $r_t$ (notice that we made use of the strong bound on $\tilde\kappa_t$ from the electron EDM)~\footnote{The complete expression including the quadratic term reads $\Gamma\p{h \rightarrow F}/\Gamma^\text{SM}\p{h \rightarrow F}=1.639-0.718\,r_t+0.079\,r_t^2 $. The last addendum is one order of magnitude smaller than the linear term and it turns out to not have any relevant impact in the results of the fit.}.

The Higgs total decay width is modified due to the NP contribution to the $t\ov{t}$-Higgs coupling present in the loop-induced process with gluons and photons in the final state:
\be
\dfrac{\Gamma_{h,\text{tot}}}{\Gamma^\text{SM}_{h,\text{tot}}} =1 + \text{BR}^\sm_{gg}\p{r_t^2 - 1}+\text{BR}^\sm_{\gamma\gamma}(0.639-0.718\,r_t) \,.
\ee

All in all, the Higgs signal strength when NP modifies only the $t\ov{t}$-Higgs coupling is then given by
\be
\mu_P^F =
\begin{cases}
\dfrac{1}{1 + \text{BR}^\sm_{gg}\p{r_t^2 - 1}+\text{BR}^\sm_{\gamma\gamma}(0.639-0.718\,r_t) } \, , \quad &\text{for } P = \text{VBF}, \text{VH} \text{ and } F \neq \gamma\gamma \\[4mm]
\dfrac{1.639-0.718\,r_t}{1 + \text{BR}^\sm_{gg}\p{r_t^2 - 1}+\text{BR}^\sm_{\gamma\gamma}(0.639-0.718\,r_t) } \, , \quad &\text{for } P = \text{VBF}, \text{VH} \text{ and } F = \gamma\gamma \\[2mm]
\dfrac{r_t^2}{1 + \text{BR}^\sm_{gg}\p{r_t^2 - 1}+\text{BR}^\sm_{\gamma\gamma}(0.639-0.718\,r_t) } \, , \quad &\text{for } P = \text{ggF}, \text{ttH}+\text{tH} \text{ and } F \neq \gamma\gamma \\[4mm]
\dfrac{r_t^2(1.639-0.718\,r_t)}{1 + \text{BR}^\sm_{gg}\p{r_t^2 - 1}+\text{BR}^\sm_{\gamma\gamma}(0.639-0.718\,r_t) } \, , \quad &\text{for }P = \text{ggF}, \text{ttH}+\text{tH} \text{ and } F = \gamma\gamma \, .
\end{cases}
\ee

We performed $\chi^2$ fits on the data listed before, obtaining constraints on the $r_\psi^2$ parameters. The results of the fit are shown at $95\%$ C.L.. Each value has been obtained switching on NP contributions only to one Yukawa coupling at a time. The results are shown in Tab.~\ref{tab:Results1Yukawa}. In the second column we show the bounds obtained from the data presented in Refs.~\cite{ATLAS:2019nkf,CMS:2018uag,ATLAS:2020fzp,CMS:2020xwi} (see Tabs.~\ref{tab:ATLASData}--\ref{tab:CMSDatamumu}).  Notice that the inclusion of the associated production data, given in the third line of Tab.~\ref{tab:ATLASDatamumu}, would require a different expression for the cross section,  $\sigma_P/\sigma^\text{SM}_P=(\sigma^\text{SM}_\text{VH}+r_t^2\sigma^\text{SM}_\text{ttH})/(\sigma^\text{SM}_\text{VH}+\sigma^\text{SM}_\text{ttH})$, and it turns out that the inclusion of this data does not change the result of the $\chi^2$ at the approximation considered.  For this reason and without any impact on the results, we decided to adopt a simplified approach not including these data. Finally, in the third column the constraints arising from the total decay width of the Higgs boson  are shown, using its experimental determination by CMS in Ref.~\cite{CMS:2019ekd} and reported in Eq.~\eqref{HiggsTotalWidthExp}.

\begin{table}[h!]
\centering{
\renewcommand{\arraystretch}{1.5}
\begin{tabular}{|c|c|c|}
\hline
\textbf{Modified Coupling} 
& $\bm{\mu_F}$ \textbf{ Ref.~\cite{ATLAS:2019nkf,CMS:2018uag,ATLAS:2020fzp,CMS:2020xwi}} 
& $\bm{\Gamma^\text{\textbf{CMS}}_{h,\text{\textbf{tot.}}}}$ \textbf{Ref.~\cite{CMS:2019ekd}}\\
\hline
\hline
$t\ov{t}h$ 		& $0.98\lesssim r_{t}^2 \lesssim 1.31$ 	&  $r_{t}^2 \lesssim 13$ \\
\hline
$c\ov{c}h$ 	& $r_{c}^2 \lesssim 2.36$ 				& $r_{c}^2 \lesssim 35$ \\
\hline
$b\ov{b}h$ 	& $ 0.73\lesssim r_{b}^2 \lesssim 1.08 $ 	& $r_{b}^2 \lesssim 2.68$ \\
\hline
$\tau\ov{\tau}h$ & $0.56 \lesssim r_{\tau}^2 \lesssim 1.40$ & $r_{\tau}^2 \lesssim 17$ \\
\hline
$\mu\ov{\mu}h$ & $0.36 \lesssim r_{\mu}^2\lesssim 1.85 $ 			 & $r_{\mu}^2 \lesssim 4600$ \\
\hline
\end{tabular}
\caption{\em $95\%$ C.L. limits on $r_{\psi}^2$ after $\chi^2$ fits to data on the Higgs signal strength from Ref.~\cite{ATLAS:2019nkf,CMS:2018uag,ATLAS:2020fzp,CMS:2020xwi} in the second column, and using the total Higgs decay width from Ref.~\cite{CMS:2019ekd} in the third column, when NP modifies  only a single fermion-Higgs coupling at a time.}
\label{tab:Results1Yukawa}
}
\end{table}

As expected, the inclusion of data on $h\to\mu\ov{\mu}$ from Refs.~\cite{ATLAS:2020fzp} and \cite{CMS:2020xwi} largely improves the bounds on $\mu\ov\mu h$. On the other hand, the data on the total Higgs decay width does not provide bounds stronger than those from the signal strengths.

\begin{figure}[H]
\centering
\includegraphics[width=0.43\textwidth]{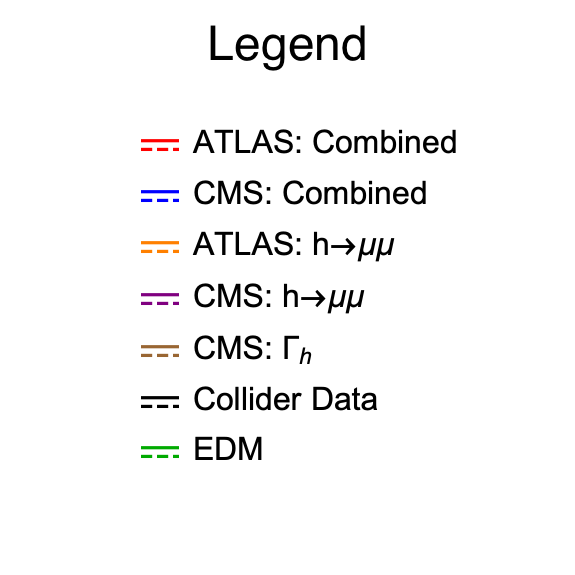}
\includegraphics[width=0.43\textwidth]{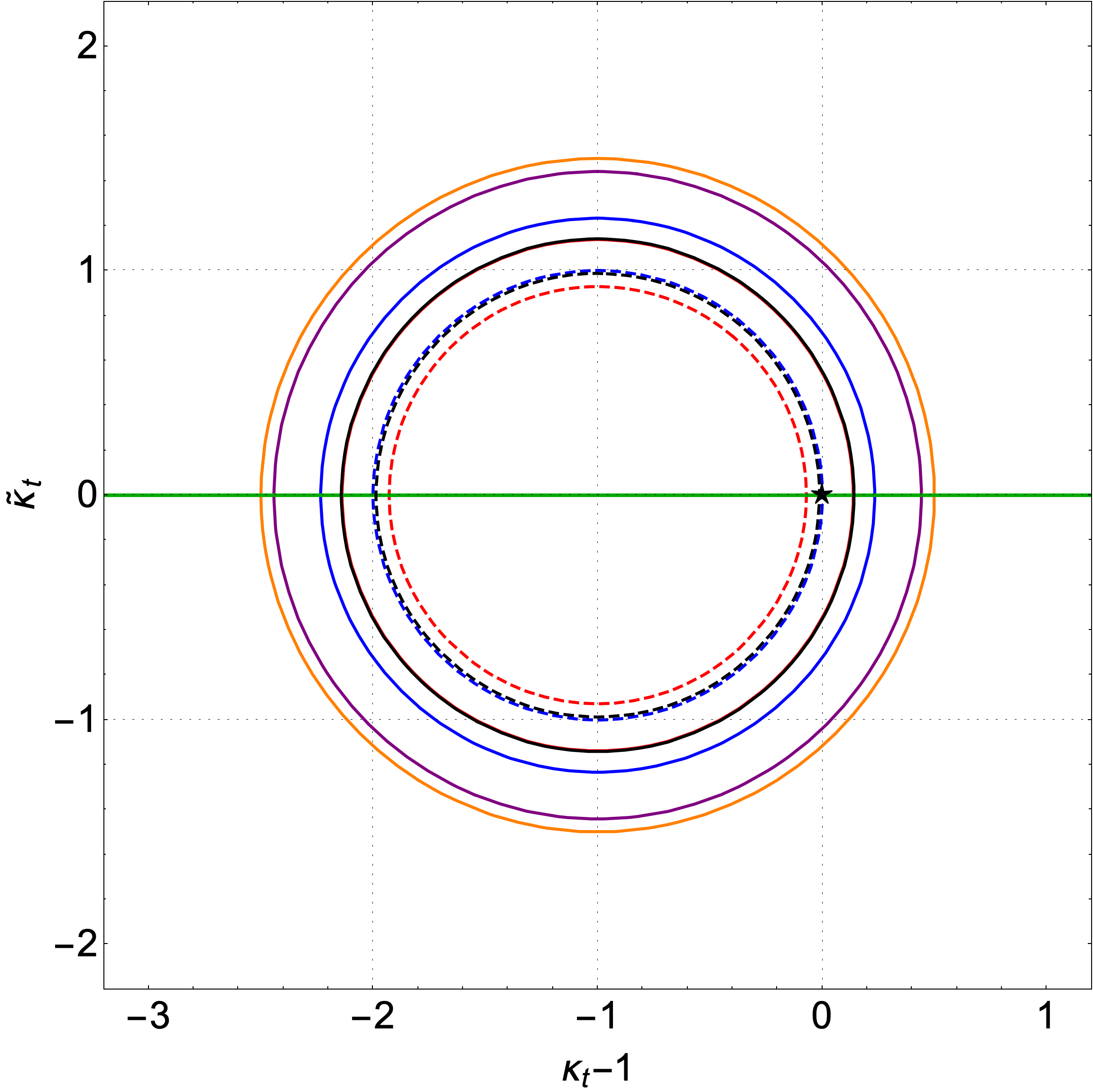}
\includegraphics[width=0.43\textwidth]{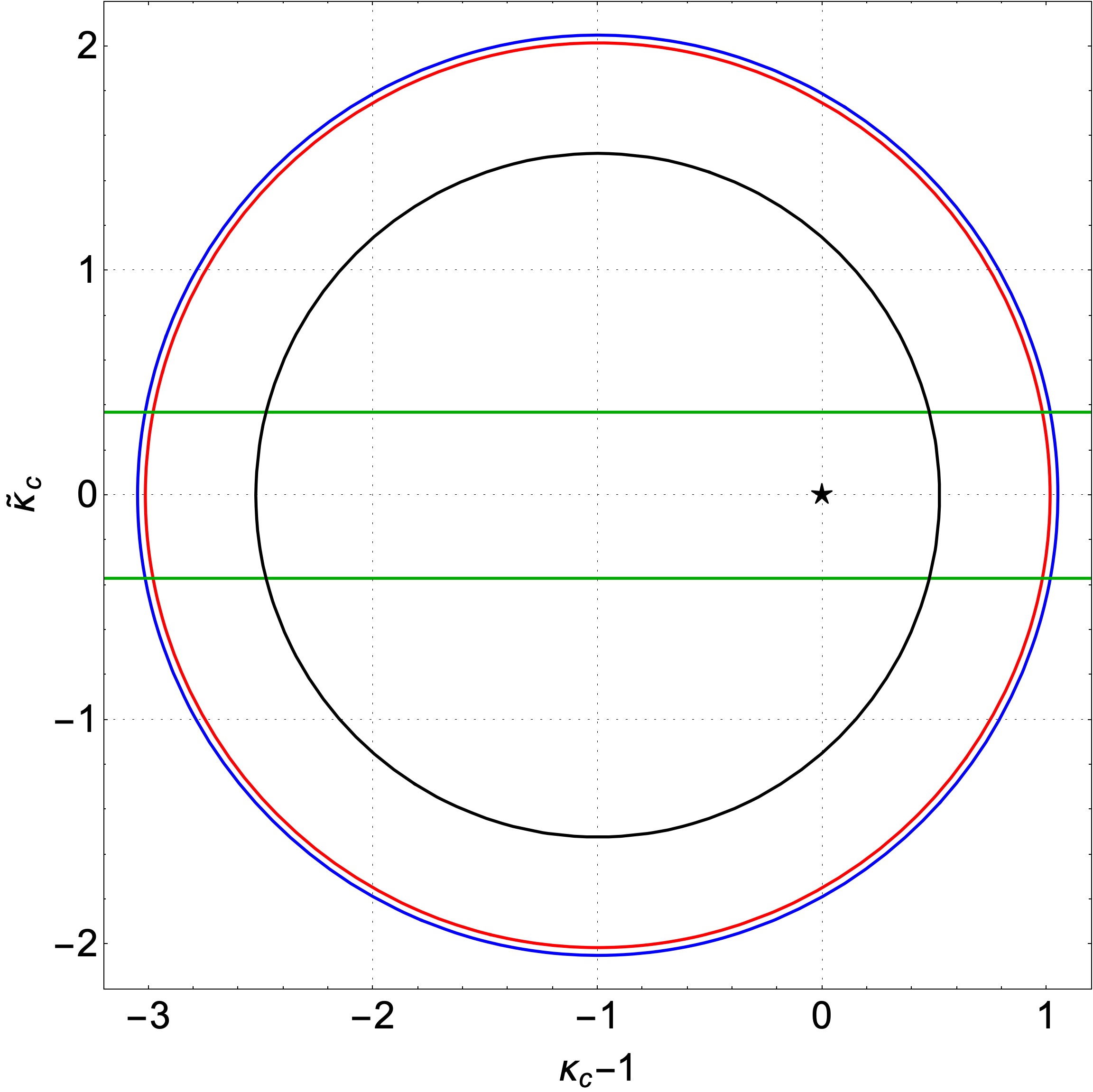}
\includegraphics[width=0.43\textwidth]{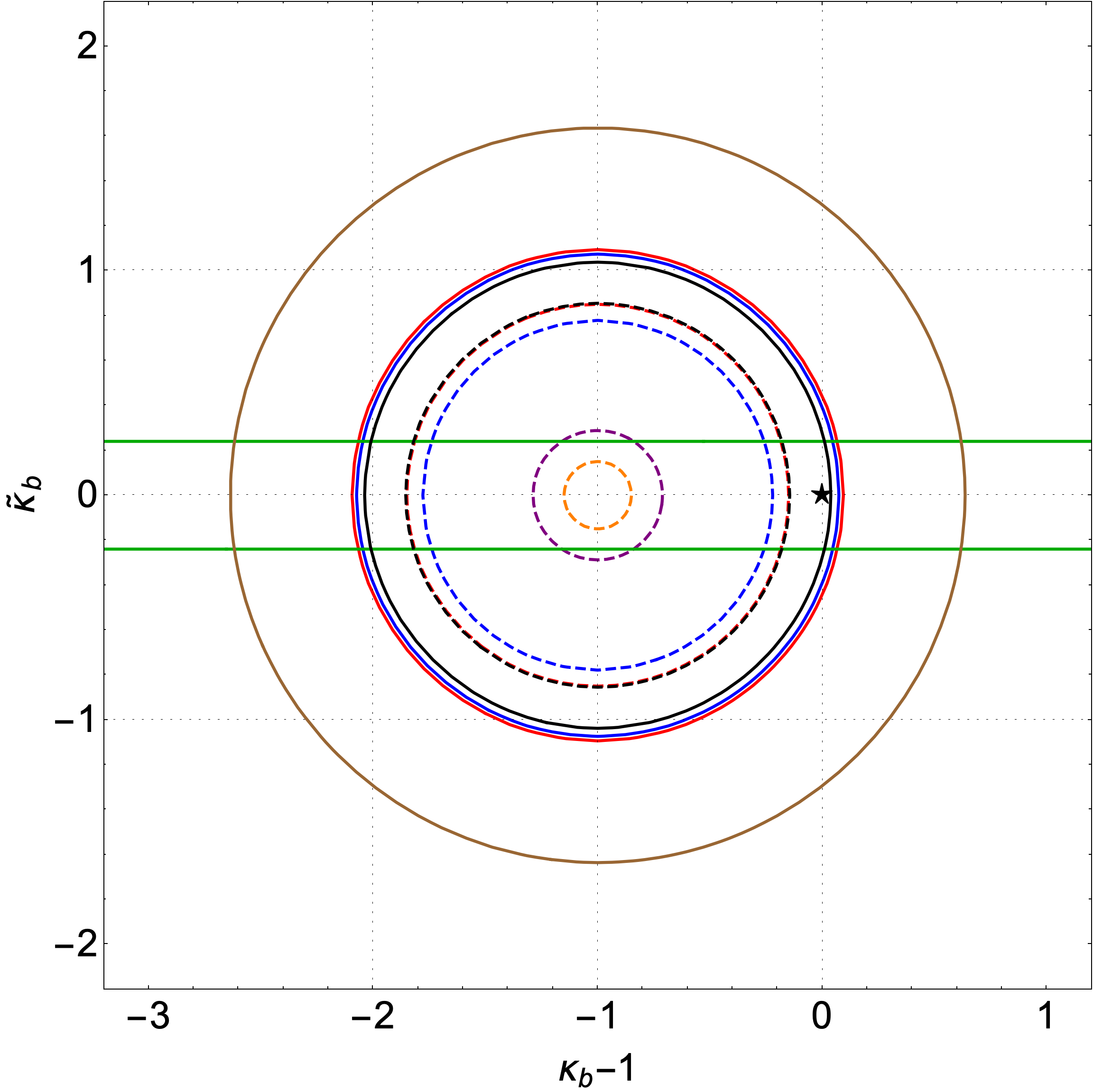}
\includegraphics[width=0.43\textwidth]{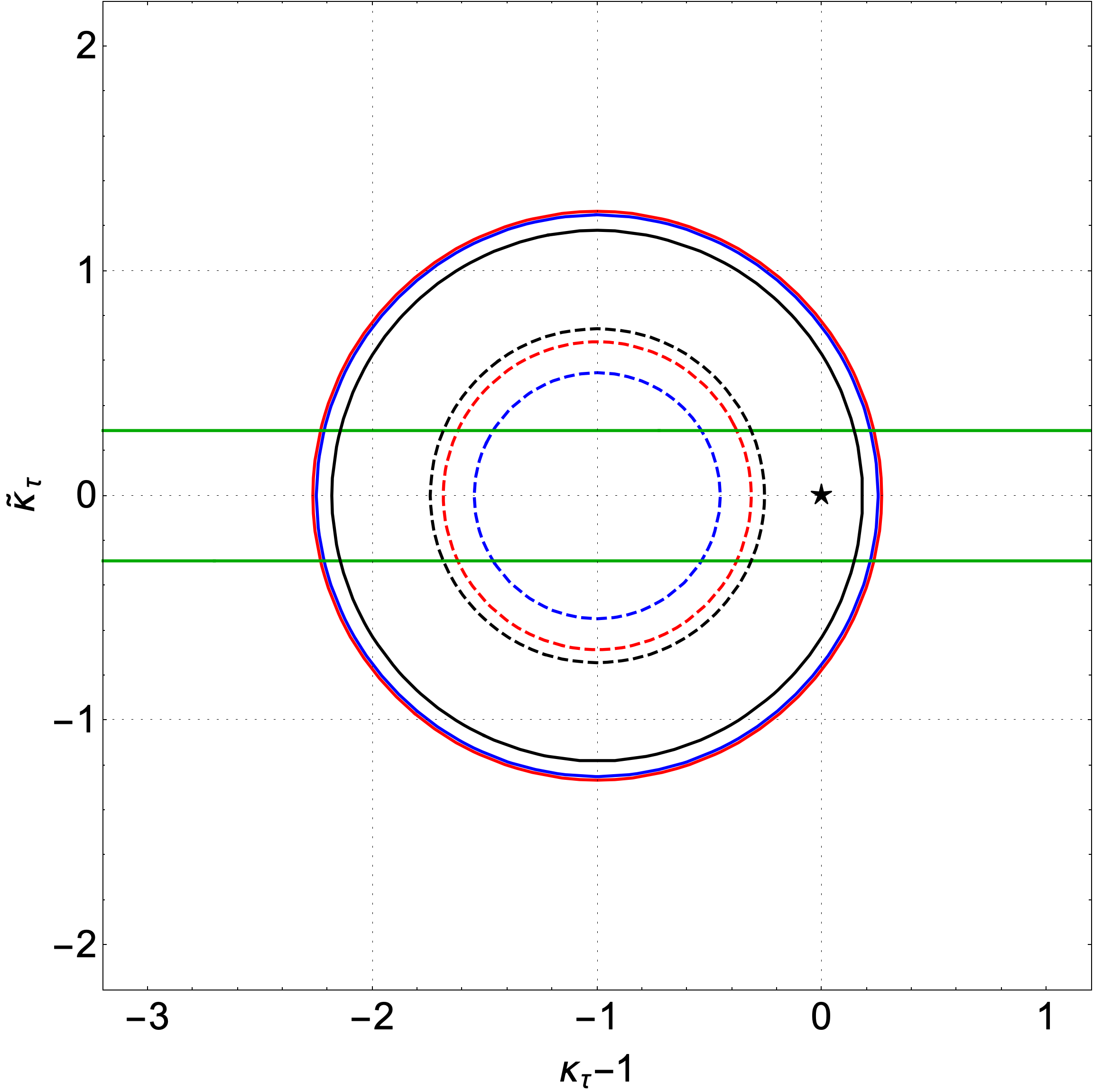}
\includegraphics[width=0.43\textwidth]{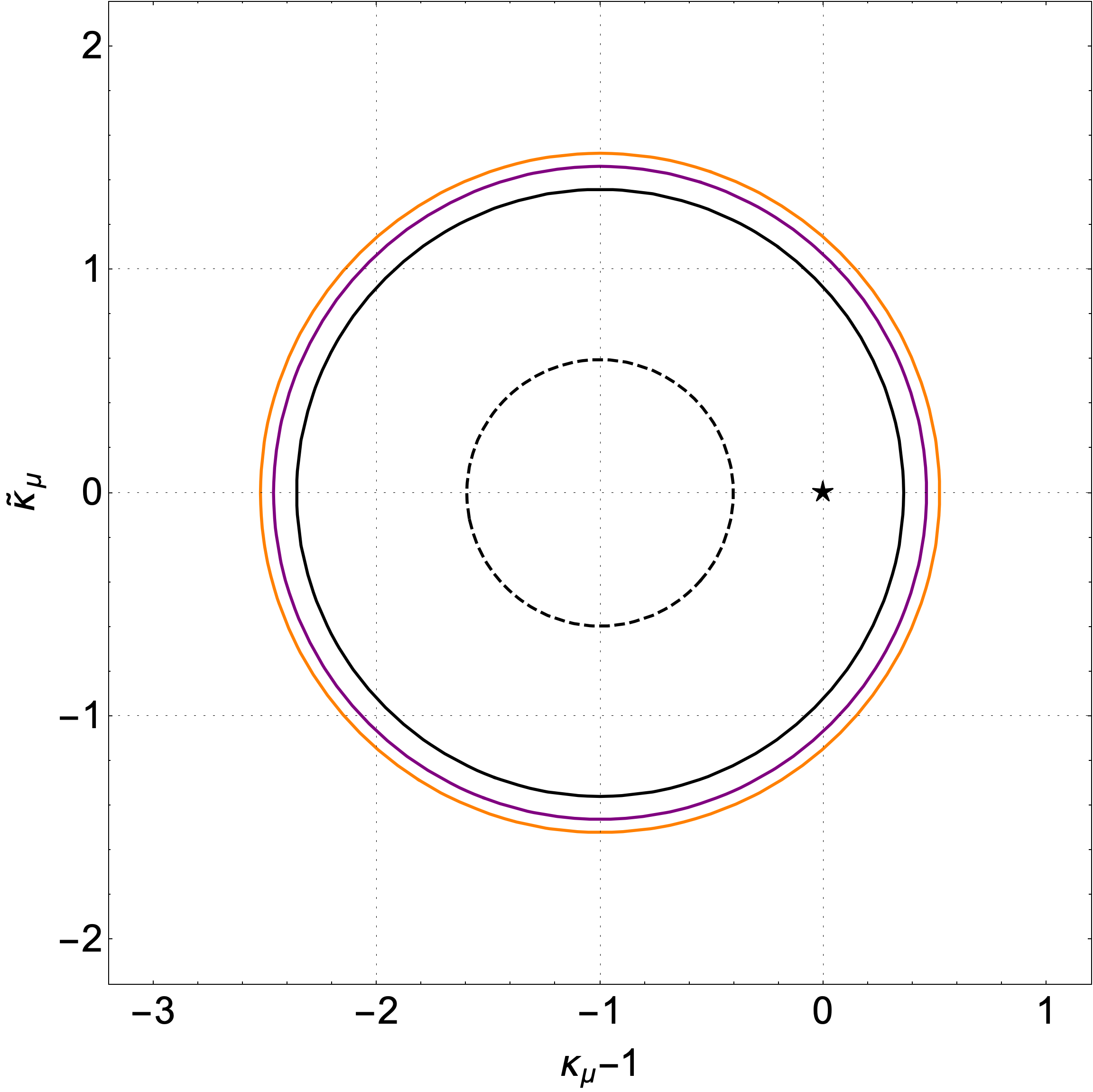}
\caption{\em Bounds in the $\p{(\kappa_\psi-1),\tilde{\kappa}_\psi}$-plane, with NP only in the single $\psi\ov{\psi}h$ coupling. These limits are derived from $\chi^2$ fits to the signal strengths and the Higgs total decay width reported in Refs.~\cite{ATLAS:2019nkf,CMS:2018uag,ATLAS:2020fzp,CMS:2020xwi} (red lines) and from electron EDM constraints from the ACME II collaboration~\cite{Andreev:2018ayy}. Continuous (dashed) lines refer to the upper (lower) bounds. The SM case ($\kappa_\psi-1= \tilde{\kappa}_\psi=0$) is represented with a star.}
\label{fig:kappa_kappatilde_plane}
\end{figure}

Fig.~\ref{fig:kappa_kappatilde_plane} shows the results of Tab.~\ref{tab:Results1Yukawa} expressed in terms of $\kappa_\psi$ and $\tilde\kappa_\psi$. The bounds obtained from data in Tab.~\ref{tab:ATLASData} are in red, from data in Tab.~\ref{tab:CMSData} are in blue, from data in Tab.~\ref{tab:ATLASDatamumu} are in orange, from data in Tab.~\ref{tab:CMSDatamumu} are in purple, from data in Eq.~\eqref{HiggsTotalWidthExp} and reported in the third column of Tab.~\ref{tab:Results1Yukawa} are in brown, while the combined results from collider data reported in the second column of of Tab.~\ref{tab:Results1Yukawa} are in black, and eventually from EDM data in Eqs.~\eqref{Boundkappatildetau1} and \eqref{BoundkappatildeCharmMuon} are in green. The star indicates the SM case with $\kappa_\psi-1= \tilde{\kappa}_\psi=0$. If a line is not present in the plot, it means that the corresponding bound is too weak and it is out of scale.

It is interesting to note that the EDM bounds can have very  strong impact. In the top case, the constraint on $\tilde\kappa_t$ is so strong that the $\kappa_t$ can only be very close to $1$ and $-1$. For the bottom and $\tau$ cases, $\kappa_{b,\tau}$ can span a slightly larger range of values around $1$ and $-1$. For charm, the absence of a lower bound on $r_c^2$ implies that $\kappa_c$ can take any value between $2$ and $-2$. Finally, for the $\mu$ plot, there is a lower (rather weak) bound on $r_\mu^2$, but the EDM one is too weak to further reduce the parameter space and therefore $\kappa_\mu$ and $\tilde\kappa_\mu$ can take almost any value between $1.5$ and $-1.5$.

\subsubsection{NP In All The Yukawa Couplings}

We can now consider the case in which all the Yukawa couplings receive contributions from NP. In this case, the cross sections for the Higgs production are: 
\begin{equation}
\frac{\sigma_P}{\sigma^\text{SM}_P} = 
\begin{cases}
1 \, , \qquad &\text{for } P = \text{VBF}, \text{VH} \\
r_t^2 \, , \qquad &\text{for } P = \text{ggF}, \text{ttH}+\text{tH} \, .
\end{cases}
\label{eq:produc_all}
\end{equation}
The Higgs partial widths are given by the following expressions:
\begin{equation}
\frac{\Gamma\p{h \rightarrow F}}{\Gamma^\text{SM}\p{h \rightarrow F}} =
\begin{cases}
1 , \qquad &\text{for } F = VV^\ast\,, \\
1.639-0.718\,r_t , \qquad &\text{for } F = \gamma\gamma \, , \\
r_b^2 , \qquad &\text{for } F = b\ov{b} \, , \\
r_\tau^2 , \qquad &\text{for } F = \tau\ov{\tau} \, , \\
r_\mu^2 , \qquad &\text{for } F = \mu\ov{\mu} \, .
\end{cases}
\label{eq:widths_all}
\end{equation}
Finally, the Higgs total width reads
\begin{equation}
\begin{split}
\frac{\Gamma_{h,\text{tot.}}}{\Gamma^\text{SM}_{h,\text{tot.}}} =& 1 + \text{BR}^\sm_{bb}\p{r_b^2 - 1} + \p{\text{BR}^\sm_{gg}+\text{BR}^\sm_{cc}}\p{r_t^2 - 1} +\\
&+\text{BR}^\sm_{\gamma\gamma}\p{0.639-0.718\,r_t}+ \text{BR}^\sm_{\tau\tau}\p{r_\tau^2 - 1} \, .
\end{split}
\label{eq:total_width_all}
\end{equation}

The signal strengths for the different processes are obtained combining properly the previous three effects. Unfortunately, the parameter space is so large that a $\chi^2$ would not be conclusive. However, assuming a flavour symmetry reduces the freedom in the parameters as already discussed in Eqs.~\eqref{ruMFV} and \eqref{ruFN}. We will present the results explicitly for the MFV case; the FN one is very similar as the only difference is the presence of the $\OO(1)$ free coefficients. 

\begin{figure}[h!]
\centering
\includegraphics[width=0.32\textwidth]{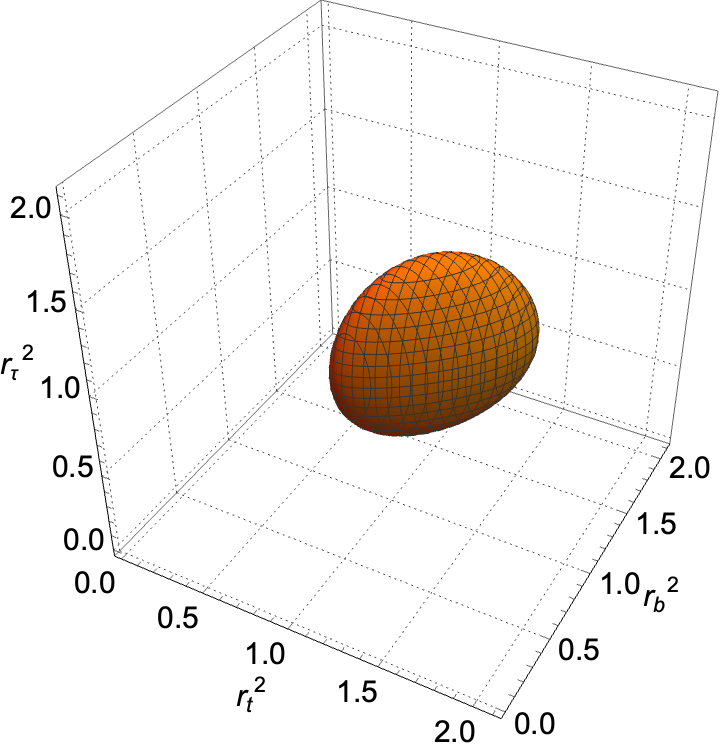}
\includegraphics[width=0.32\textwidth]{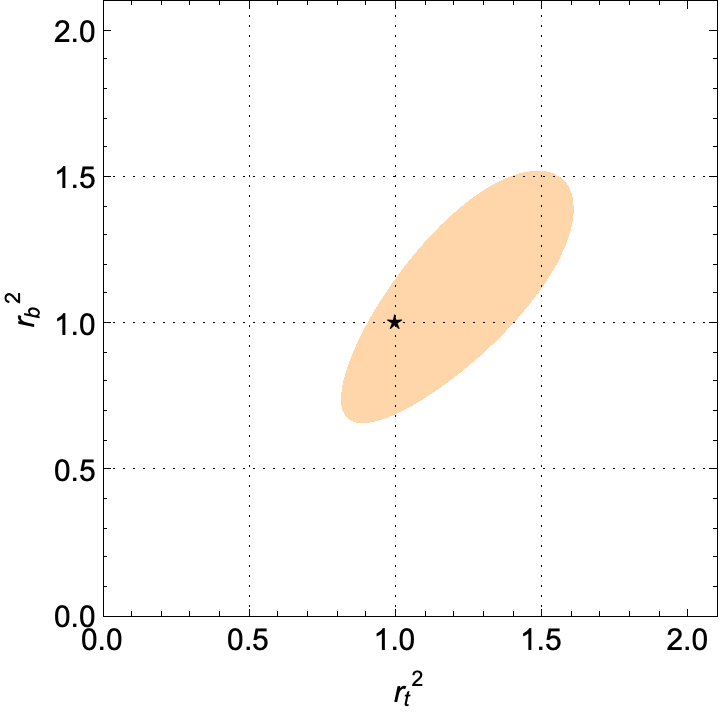}
\includegraphics[width=0.32\textwidth]{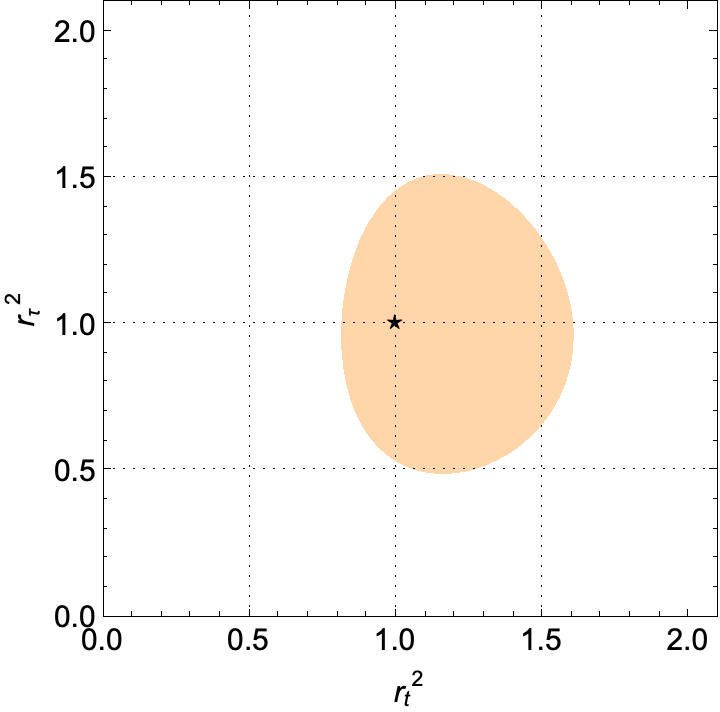}
\includegraphics[width=0.32\textwidth]{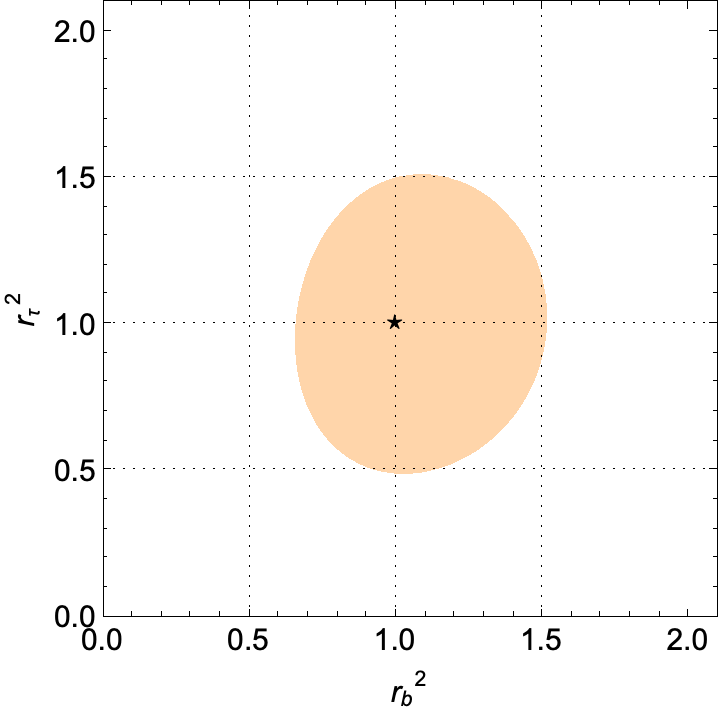}
\includegraphics[width=0.39\textwidth]{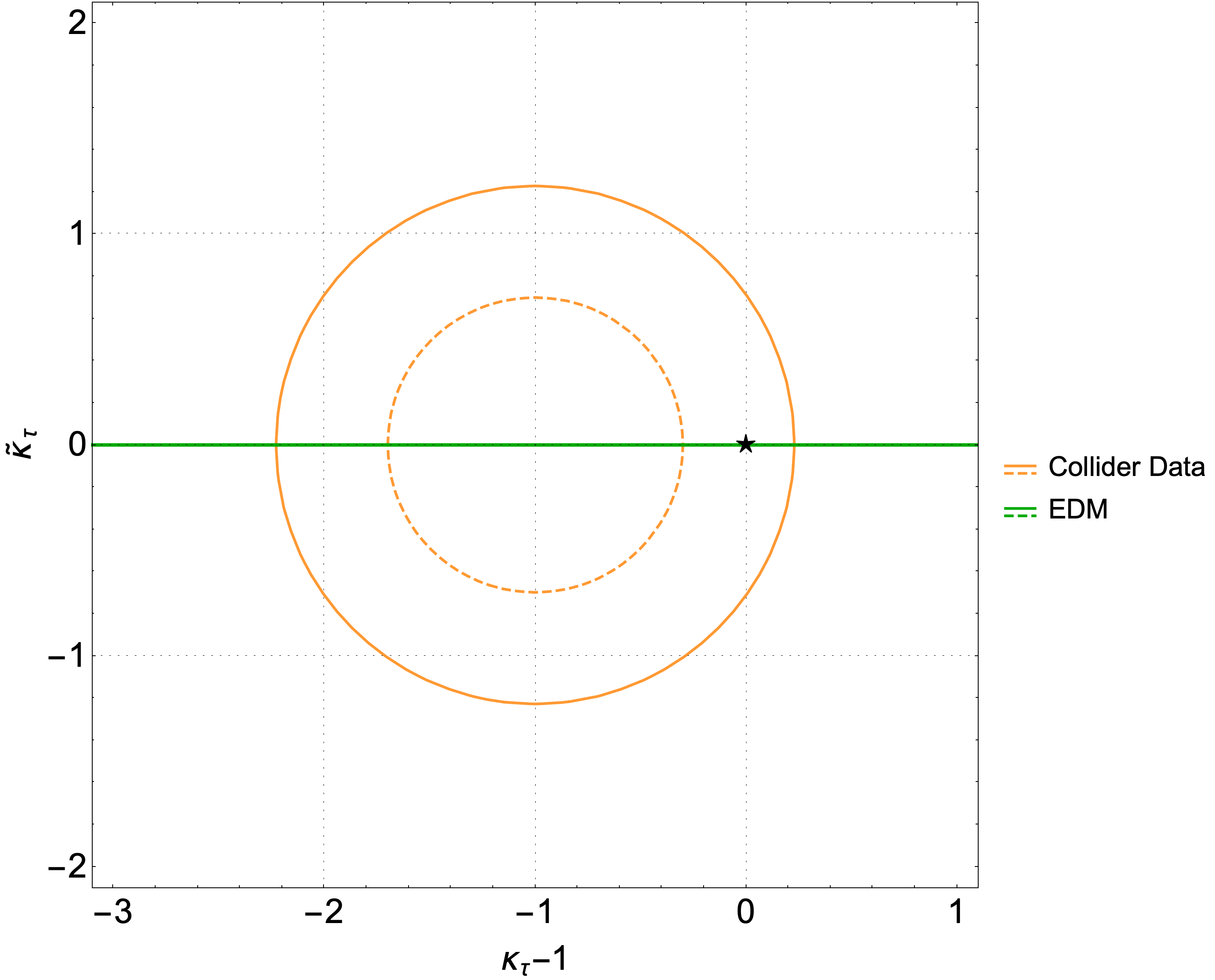}
\caption{\em Bounds in the $r_\psi^2$-$r_{\psi}^{\prime 2}$ planes, with NP contributions to all the Yukawa couplings. The MFV case is shown  and the FN one is  very similar. The plot on top-left  is the main result of the $\chi^2$ fit, and the following three ones are the projections in two-dimensions. The last plot shows the $\kappa_\tau-\tilde\kappa_\tau$ parameter space: in orange the result of the global $\chi^2$ fit and in green the EDM bound as reported in Eq.~\eqref{Boundkappatildetau2}. The SM case ($r^2_\psi=1$ or $\kappa_\tau-1= \tilde{\kappa}_\tau=0$) is represented with a star.}
\label{fig:kappa_kappatilde_3}
\end{figure}

In Fig.~\ref{fig:kappa_kappatilde_3} we plot the final result of the $\chi^2$ fit on the full parameter space, where $r_u^2=r_c^2=r_t^2$, $r_d^2=r_s^2=r_b^2$ and $r_e^2=r_\mu^2=r_\tau^2$. Moreover, we show the projections in two dimensions that allow to extract the bounds on the different parameters. The data used for the $\chi^2$ fit  are those from Refs.~\cite{ATLAS:2019nkf,CMS:2018uag,ATLAS:2020fzp,CMS:2020xwi}: as for the previous analysis whose results are shown in Tab.~\ref{tab:Results1Yukawa}, the only data that have not been used are the ones from the last category of events in Tab.~\ref{tab:ATLASDatamumu} denoted as VH$+t\ov{t}$H. Moreover, when analysing the data from Ref.~\cite{ATLAS:2019nkf}, we implemented the correlations provided in the original paper; on the other side, we assumed that data from different papers are uncorrelated. In the last plot of Fig.~\ref{fig:kappa_kappatilde_3}, we show the $\kappa_\tau-\tilde\kappa_\tau$ parameter space: while the ones for top and bottom are very similar to the ones in Fig.~\ref{fig:kappa_kappatilde_plane}, for the $\tau$ there is a  clear change represented by the EDM bound in green. For the MFV case, and very similarly for the FN one, the induced EDM bound is very strong and comparable to the one on the top, as discussed in Eq.~\eqref{Boundkappatildetau2}.

\begin{table}[h!]
\centering{
\renewcommand{\arraystretch}{1.7}
\begin{tabular}{|c|}
\hline
\textbf{Bounds from Refs.~\cite{ATLAS:2019nkf,ATLAS:2020fzp,CMS:2018uag,CMS:2020xwi}} \\
\hline
\hline
$0.82 \lesssim r_{t,c,u}^2 \lesssim 1.61$ \\
 $0.67 \lesssim r_{b,s,d}^2 \lesssim 1.52$ \\
 $0.49 \lesssim r_{\tau,\mu, e}^2  \lesssim 1.51$\\
\hline
\end{tabular}
\caption{\em $95\%$ C.L. limits on $r_{\psi}^2$ after $\chi^2$ fits obtained on the data in Refs.~\cite{ATLAS:2019nkf,ATLAS:2020fzp,CMS:2018uag,CMS:2020xwi}, when NP contributes to all the Yukawa couplings. These results holds for the MFV case, while the FN one is likely to be very similar.}
\label{tab:all_sectors}}
\end{table}

In tab.~\ref{tab:all_sectors}, we report the bounds for $r_{t,c,u}^2$, $r_{b,s,d}^2$ and $ r_{\tau,\mu, e}^2$ from Fig.~\ref{fig:kappa_kappatilde_3}. Comparing these results with those in Tab.~\ref{tab:Results1Yukawa}, we can see that the bounds are slightly weaker. On the other side, the $r_\psi^2$ of the lighter generations are also constrained: for the $c\ov{c}h$ and $\mu\ov\mu h$ couplings, we can see that the inferred bounds are stronger than those we could put in the absence of any symmetry; moreover, we can find constraints on $u\ov{u}h$, $d\ov{d}h$, $s\ov{s}h$ and $e\ov{e}h$.

The idea  of constraining theoretically the diagonal light generation Yukawas has been proposed in Ref.~\cite{Goertz:2014qia,Perez:2015aoa,Perez:2015lra}. In that paper, they are linked to off-diagonal Higgs-quark couplings contributing
at the tree level  to the $K$, $B$ and $D$ meson oscillations. 
The misalignment between the quark mass and the Yukawa matrices is introduces by  adding dim 6 operators with anarchical flavour structure to the SM Yukawa coupling. The diagonal and off-diagonal couplings are related by arbitrary coefficients that are randomly generated within a specific numerical interval in a numerical analysis. The  bounds of 
Ref.~\cite{Goertz:2014qia} parametrised in terms of $r_d$ read:
\be
0.16\lesssim r_d^2\lesssim 2.89\,.
\ee
Comparing this result to the one in Tab.~\ref{tab:all_sectors} we can see that, once a flavour symmetry is acting in the SM Lagrangian, present colliders bounds on $\hat Y_d$ are already stronger than those from flavour observables.\\

Before moving to discuss NP in the Higgs   boson Yukawa couplings from the perspective of  flavour changing processes, we summarise  the results of this section in terms of the  lower bounds  of the cut-off scales $\Lambda_{q,\ell}$ emerging from our analyses based on the assumption of an underlying  flavour symmetry.  For the sake of a transparent presentation, one can distinguish two cases: large CP violating phases of the effective Yukawa couplings and approximately CP conserving scenario. The bounds on the cut-off scales $\Lambda_{q,\ell}$  following from the electron EDM and the collider Higgs boson production and flavour  conserving decays  are then  summarised in  Tab.~\ref{tab:SummaryBounds}. These values strictly holds for the MFV case and also, up  to $\OO(1)$ coefficients,  in the FN case. 

\begin{table}[h]
\centering{
\renewcommand{\arraystretch}{1.7}
\begin{tabular}{|c||c|c|}
\hline
									& {\bf Conditions on $\theta_{f,ii}$} 				&  {\bf Bound} \\
\hline
\hline
\multirow{2}{*}{\bf EDM }					& $\sin\theta_{u,33}=1$ 						& $\Lambda_q\gtrsim 7.4\TeV$ \\
									& $\sin\theta_{e,11}=1=\sin\theta_{e,33}$ 			& $\Lambda_\ell\gtrsim 6.0\TeV $ \\
\hline
\multirow{2}{*}{\bf Collider (diag couplings)}	& $\sin\theta_{u,33}=0$						& $\Lambda_q\gtrsim 0.8\TeV$ \\
									& $\sin\theta_{e,11}=0=\sin\theta_{e,33}$		& $\Lambda_\ell\gtrsim 0.5\TeV$ \\
\hline
\end{tabular}
\caption{\em Summary of the bounds on $\Lambda_{q,\ell}$ due to EDMs and collider Higgs boson production data. In the second column, the corresponding conditions on the phases $\theta_{f,ii}$ are also reported.}
\label{tab:SummaryBounds}}
\end{table}

The results in the first line are those given in Eq.\eqref{BoundsLambdaEDMs}. The values of $\Lambda_{q,\ell}$ explain the
experimental bounds on $\tilde{\kappa}_{t,e,\tau}$, with the corresponding $\sin\theta=1$ in Eq.~\eqref{KfKf}. 
The results in the second line follow from Eqs.~\eqref{rpsiDef} and \eqref{ruFN}: the strong experimental EDM bounds on the $\tilde{\kappa}_{t,e,\tau}$ imply  $r_\psi\approx\kappa_\psi$ for $\psi=t,\,e,\,\tau$. Moreover, we have assumed that in Eq.~\eqref{KfKf}, $\sin\theta=0$ for $\psi=t,\,e,\,\tau$. It follows from our fits that the role of flavour symmetry in obtaining those bounds appears only at the quantitative rather than qualitative level, namely,  without an underlying flavour symmetry  the bounds become flavour dependent but in the same bulk part. The only significant difference is for the $\tau$: the EDM bound on $\Lambda_\ell$ would be 13 times weaker without an underlying flavour symmetry and assuming CP violating NP only in the $\tau\ov\tau h$ coupling.
On the contrary, an underlying flavour symmetry has strong implications for the discussion in the next section.

\section{Bounds from Flavour Changing Processes}
\label{sec:OffDiag}

In this section we use the upper bounds on the effective flavour non-diagonal Yukawa couplings derived from tree-level Higgs exchange contributions~\cite{Harnik:2012pb} to various $\Delta F=2$ processes and radiative decays of quarks and leptons and calculate the lower bounds on the NP scales, under the assumption that the corresponding Wilson coefficients are controlled by a FN symmetry.  The observables considered in this section may receive NP contributions different from those we  have considered  so far and that can also be described at low-energy through four fermion interactions. For this reason, the results obtained in the following should be considered as a conservative estimation.

We use previous results presented in Ref.~\cite{Blankenburg:2012ex}, updating the bounds when new, more precise experimental values are present, where the  EFT approach has been adopted, with  the physical Higgs integrated out. In that article, observables in both quark and lepton sectors have been taken into account. In Tab.~\ref{tab:QboundsFinal}, we list the combination of the effective Yukawa couplings $\hat{y}_{\psi\psi'}$, defined as the entries of $\hat Y$ in Eq.~\eqref{hatYf}, for the quark sector: in the second column we report the bound on the absolute value of the given combination, and  in the fourth column on the imaginary part of that combination. Notice that the notation used in Ref.~\cite{Blankenburg:2012ex} and in these tables is different: the $c_{\psi\psi'}$ coefficients of Ref.~\cite{Blankenburg:2012ex} are here the effective Yukawas $\hat{y}_{\psi\psi'}$ and the relation that links the two quantities is very simple, $\hat{y}_{\psi\psi'}=\sqrt2 c_{\psi\psi'}$. 

\begin{table}[h!]
\centering
\begin{tabular}{|c||c|c||c|c|}
\hline
&&&&\\[-3mm]
{\bf Eff. Coupl.} 
& {\bf Bound on $|y_{\text{eff.}}|$ } 
& {\bf $\Lambda_q$ [$\TeV$] } 
& {\bf Bound on $|\Im(y_{\text{eff.}})|$ } 
& {\bf$\Lambda_q$ [$\TeV$]} \\ [1mm]
\hline
\hline
&&&&\\[-3mm]
$\hat{y}_{sd} \hat{y}_{ds}^*$ 
& $2.2 \times 10^{-10}$
& $0.7$
& $8.2 \times 10^{-13}$
&$3 $\\
$\hat{y}_{cu} \hat{y}_{uc}^*$ 
& $1.8 \times 10^{-9}$
& $0.7$
& $3.4\times 10^{-10}$
&$1$\\
$\hat{y}_{bd} \hat{y}_{db}^*$ 
& $1.8\times 10^{-8}$
& $0.6$
& $5.4 \times 10^{-9}$
&$0.8$\\
$\hat{y}_{bs} \hat{y}_{sb}^*$ 
& $4.0\times 10^{-7}$
& $0.6$
& $4.0 \times 10^{-7}$
&$0.6$\\[1mm]
\hline
&&&&\\[-3mm]
$\hat{y}_{ds}^2$ 
& $4.4 \times 10^{-10}$
& $0.6$
& $1.6\times 10^{-12}$
&$ 2$\\
$\hat{y}_{sd}^2$
& $4.4 \times 10^{-10}$
& $0.6$
& $1.6\times 10^{-12}$
& $2$\\
$\hat{y}_{uc}^2$ 
& $2.8 \times 10^{-9}$
& $1$
& $5.0\times 10^{-10}$
&$ 2$\\
$\hat{y}_{cu}^2$ 
& $2.8 \times 10^{-9}$
& $0.2$
& $5.0\times 10^{-10}$
&$0.4$\\
$\hat{y}_{db}^2$ 
& $2.0 \times 10^{-8}$
& $0.4$
& $6.0 \times 10^{-9}$
&$0.5$\\
$\hat{y}_{bd}^2$ 
& $2.0 \times 10^{-8}$
& $1$
& $6.0 \times 10^{-9}$
&$1$\\
$\hat{y}_{sb}^2$ 
& $4.4 \times 10^{-7}$
& $0.3$ 
& $4.4 \times 10^{-7}$
&$ 0.3$\\
$\hat{y}_{bs}^2$ 
& $4.4 \times 10^{-7}$
& $1$
& $4.4 \times 10^{-7}$
&$1$\\[1mm]
\hline
\end{tabular}
    \caption{\em FN lower bounds on the NP cut-off scale $\Lambda_q$ in TeV following from the off-diagonal effective Yukawa couplings in the quark sector. The bounds in the first half of the table are independent of the FN charges, and the others are obtained taking $\epsilon=0.23$ and with the charge assignments  given in Tab.~\ref{tab:FNscen}. The bounds are extracted from Ref.~\cite{Blankenburg:2012ex}. The FN $\mathcal{O}(1)$ coefficients are taken equal to 1.}
 \label{tab:QboundsFinal}
\end{table}

Those bounds, taken at the face value, suggest a strong power of flavour observables in searching for NP in the Yukawa couplings (see Fig.~5.1 in Ref.~\cite{EuropeanStrategyforParticlePhysicsPreparatoryGroup:2019qin}). Indeed, assuming the structure of the effective couplings to be $\OO(1)v^4/\Lambda_q^4$, those observables would be sensitive to the scales as high as $\Lambda_q\approx (60-300)\TeV$, with the lower value given by the  bounds on the absolute value of the effective Yukawas $\hat{y}_{sd} \hat{y}_{ds}^*$ and the upper one by its imaginary part. The  picture is, however, strikingly different if the Wilson coefficients of the dim 6 operators contributing to the Yukawa couplings respect an underlying flavour symmetry.  This clearly emerges from the third and fifth columns of Tab.~\ref{tab:QboundsFinal}, where we show the corresponding lower bounds on the NP scales when the Wilson coefficients respect the FN symmetry. The bounds given in the third column follow from the limits on the absolute values of the effective coefficients and  are independent of the phases $\theta_{f,ij}$. The  bounds  in the fifth column follow from the limits on the imaginary parts, once we assume the corresponding $\sin\theta_{f,ij}=1$. In all cases, the free $\mathcal{O}(1)$ coefficients are taken strictly equal to 1.

The obtained values for the lower  bounds on NP scales are subject to three main sources of uncertainty: the first one is due to the $\OO(1)$ coefficients typical of the FN context; the second is associated to the experimental errors on the lightest quark masses; the third one is linked to the running between the scale $\Lambda_F$ down to the scale at which the different observables are computed~\footnote{Although the exact dependence of the observables on the Lagrangian Wilson coefficients is typically affected by the running, assuming the absence of \textit{ad hoc} cancellations between different contributions, the numerical results remain in the same ballpark.}. Therefore, they should be considered as approximate, qualitative results.

The bounds in the upper part of the table are independent from the FN charge assignment, because for the corresponding $\hat y$ combinations it is possible to re-express the specific FN charge dependence in terms of quark masses. This does not occurs for the bounds in the lower part of the table for which the charges in Tab.~\ref{tab:FNscen} have been taken. As can be seen, for most of the combinations of effective Yukawas, the values of the scale $\Lambda_q$ following from the experimental limits on their absolute values are very similar to the collider value $\Lambda_q=0.8$ TeV. The exceptions are $\hat{y}^2_{uc}$, $\hat{y}^2_{bs}$, $\hat{y}^2_{bd}$ which give
\be
\Lambda_q\gtrsim1\TeV\,,
\ee
still very close to the collider limits. This is the effect of the underlying flavour symmetry.

Concerning the imaginary parts of the effective Yukawa combinations and assuming maximal phases $\sin\theta=1$,  the obtained bounds  on  several  of $\Lambda_q$'s are in a couple of TeV range.  The stronger one reads
\be
\Lambda_q\gtrsim3\TeV\qquad\qquad (\Im(\hat{y}_{ds}\hat{y}^*_{ds}) \text{ from $K^0$ meson oscillation}).
\ee
These values are different quantitatively from the one extract from colliders data, but are still orders of magnitude below the values of $\Lambda_q$ obtained when Wilson coefficients are not controlled by a flavour symmetry. It is interesting to notice that neither the collider nor the flavour data are competitive with those from the electron EDM, from which we can extract the strongest bounds, see Tab.~\ref{tab:SummaryBounds}. If all the phases are maximal, including $\theta_{u,33}$, then $\Lambda_q\gtrsim7.4\TeV$. On the other hand, if for any reason $\sin\theta_{u,33}=0$ while the non-diagonal couplings have maximal CP violating phases, then the EDM bounds would not apply and flavour data would provide the strongest bounds, reported in the last column of Tab~\ref{tab:QboundsFinal}.

In Tab.~\ref{tab:boundsFinal}, we discuss leptons: the last three columns correspond to the three leptonic FN models. The limits in the upper part of the table are independent from the FN charges and the bounds are in common to all the models: also in this case, it is possible to express the FN charge dependence on the corresponding effective Yukawa combinations in terms of lepton masses. This is not the case for the bounds in the lower part of the table. 

\begin{table}[h!]
\centering
\begin{tabular}{|c||c|c|c|c|}
\hline
&& \multicolumn{3}{|c|}{}\\[-3mm]
\multirow{2}{*}{\bf Eff. Coupl.} & \multirow{2}{*}{\bf Bound} &  \multicolumn{3}{|c|}{\bf $\Lambda_\ell$ [$\TeV$]}\\[1mm]
\cline{3-5}&&&&\\[-4mm]
&&$A$ & $A_{\mu\tau}$ & $H$ \\ [1mm]
\hline
\hline
&&&&\\[-3mm]
$|\hat{y}_{e\tau}\hat{y}_{\tau e}|$ &  $2.2 \times 10^{-2}$\cite{Blankenburg:2012ex} &
$8.\times 10^{-3}$ & $=$ & $=$ \\
$|\Re(\hat{y}_{e\tau}\hat{y}_{\tau e})|$ &  $1.4 \times 10^{-4}$\cite{Zyla:2020zbs}& 
$3\times 10^{-2}$ & $=$ & $=$ \\
$|\Im(\hat{y}_{e\tau}\hat{y}_{\tau e})|$ & $1.1 \times 10^{-10}$ \cite{Zyla:2020zbs} & 
$1$ & $=$ & $=$ \\
$|\hat{y}_{e\mu}\hat{y}_{\mu e}|$ & $3.6 \times 10^{-1}$\cite{Blankenburg:2012ex} &
$2\times 10^{-3}$ & $=$ & $=$ \\
$|\Re(\hat{y}_{e\mu}\hat{y}_{\mu e})|$ & $1.4 \times 10^{-3}$\cite{Zyla:2020zbs} & 
$8\times 10^{-3}$ & $=$ & $=$ \\
$|\Im(\hat{y}_{e\mu}\hat{y}_{\mu e})|$ & $1.1 \times 10^{-9}$\cite{Zyla:2020zbs}& 
$0.3$ & $=$ & $=$ \\
$|\hat{y}_{\mu\tau}\hat{y}_{\tau\mu}|$  & $4.0$\cite{Blankenburg:2012ex} &
$9\times 10^{-3}$ & $=$ & $=$ \\
$|\Re(\hat{y}_{\mu\tau}\hat{y}_{\tau\mu})|$  & $2.5 \times 10^{-3}$\cite{Zyla:2020zbs}& 
$5\times 10^{-2}$ & $=$ & $=$ \\
$|\Im(\hat{y}_{\mu\tau}\hat{y}_{\tau\mu})|$ & $1.2$\cite{Blankenburg:2012ex} & 
$1\times 10^{-2}$ & $=$ & $=$ \\[1mm]
\hline 
&&&&\\[-3mm]
$|\hat{y}_{e\tau}\hat{y}_{\tau \mu}|$ & $6 \times 10^{-8}$ \cite{Zyla:2020zbs}& 
$0.8$&$0.5$&$0.5$ \\
$|\hat{y}_{\tau e }\hat{y}_{\mu \tau }|$ & $6\times 10^{-8}$ \cite{Zyla:2020zbs} & 
$0.2$&$0.3$&$0.3$\\
$|\hat{y}_{\mu\tau}|^2$  & $2.5\times 10^{-6}$ \cite{CMS:2021rsq}&
$0.6$&$0.6$&$0.3$ \\
$|\hat{y}_{\tau\mu}|$  & $2.5 \times 10^{-6}$\cite{CMS:2021rsq}& 
$0.2$&$0.2$&$0.3$ \\
$|\hat{y}_{e\tau}|^2$ & $3.6 \times 10^{-6}$ \cite{CMS:2021rsq}& 
$0.6$&$0.3$&$0.1$ \\
$|\hat{y}_{\tau e}|^2$ & $3.6 \times 10^{-6}$\cite{CMS:2021rsq}& 
$1\times 10^{-2}$&$2\times 10^{-2}$&$4\times 10^{-2}$ \\
$|\hat{y}_{\mu e}|^2$  & $3.5 \times 10^{-12}$\cite{Zyla:2020zbs}& 
$0.3$&$0.6$&$0.6$\\
$|\hat{y}_{e\mu}|^2$ & $3.5\times 10^{-12}$ \cite{Zyla:2020zbs}& 
$4$&$2$&$2$ \\
$|\hat{y}_{\mu e}\hat{y}_{e\tau }^*|$  & $1.8 \times 10^{-4}$\cite{Blankenburg:2012ex}
& $3\times 10^{-2}$&$3\times 10^{-2}$&$2\times 10^{-2}$ \\
$|\hat{y}_{\mu e}\hat{y}_{\tau e}|$ & $1.8 \times 10^{-4}$\cite{Blankenburg:2012ex}
&  $4\times 10^{-3}$&$8\times 10^{-3}$&$1\times 10^{-2}$ \\
$|\hat{y}_{e\mu}^* \hat{y}_{e\tau }^*|$  & $1.8 \times 10^{-4}$\cite{Blankenburg:2012ex}
& $0.1$&$5\times 10^{-2}$&$4\times 10^{-2}$ \\
$|\hat{y}_{e\mu}^* \hat{y}_{\tau e}|$  & $1.8 \times 10^{-4}$\cite{Blankenburg:2012ex}
& $1\times 10^{-2}$&$1\times 10^{-2}$&$2\times 10^{-2}$ \\
$|\hat{y}_{e\mu}\hat{y}_{\mu\tau}^*|$  & $2.0 \times 10^{-4}$\cite{Blankenburg:2012ex}
& $0.1$&$7\times 10^{-2}$&$5\times 10^{-2}$ \\
$|\hat{y}_{e\mu}\hat{y}_{\tau \mu}|$   & $2.0 \times 10^{-4}$\cite{Blankenburg:2012ex}
& $5\times 10^{-2}$&$4\times 10^{-2}$&$5\times 10^{-2}$\\
$|\hat{y}_{\mu e}^*\hat{y}_{\mu\tau}^*|$  & $2.0 \times 10^{-4}$\cite{Blankenburg:2012ex}
&$3\times 10^{-2}$&$4\times 10^{-2}$&$3\times 10^{-2}$\\
$|\hat{y}_{\mu e}^*\hat{y}_{\tau\mu}|$  & $2.0 \times 10^{-4}$\cite{Blankenburg:2012ex}
& $1\times 10^{-2}$&$2\times 10^{-2}$&$3\times 10^{-2}$ \\[1mm]
\hline
\end{tabular}
    \caption{\em FN lower bounds on the NP cut-off scale $\Lambda_\ell$ in TeV following from the off-diagonal effective Yukawa couplings in the leptonic sector.  Most of the bounds are taken from Ref.~\cite{Blankenburg:2012ex}, except those that show explicitly a different reference: in these cases, the experimental bound has improved  as compared to the one reported in Ref.~\cite{Blankenburg:2012ex}. The bounds in the first half of the table are independent of the FN charges, while the others are obtained taking $\epsilon=0.23$ and with the charge assignments  given in Tab.~\ref{tab:FNscen}.  ``='' indicates that the bounds in the $A_{\mu\tau}$ and $H$ models are the same as those in the $A$ model.}
    \label{tab:boundsFinal}
\end{table}

For most of the combinations of effective Yukawas for leptons, the lower bounds on the scale $\Lambda_\ell$  are  (much) lower than $\Lambda_\ell = 0.5\TeV$ obtained from the collider   data.  The main exception is $|\hat{y}_{e\mu}|^2$ that is linked to the muon radiative decay.  Barring  large cancellations between  free $\OO(1)$ coefficients, we obtain
\be
\begin{cases}
\Lambda_\ell=4 \TeV &\qquad\text{for the $A$ Model}\\
\Lambda_\ell=2 \TeV &\qquad\text{for the $A_{\mu\tau}$ and $H$ Models}\\
\end{cases}
\qquad (|\hat{y}^2_{e\mu}| \text{ from $\mu\to e\gamma$ decay})\,,
\ee
as the lower bound on the NP scale (independently of the phases of the individual Yukawa couplings). Similarly to the quark case, these values for $\Lambda_\ell$ should be contrasted with the value $\Lambda_\ell=300\TeV$ for the NP scale if no flavour symmetry is invoked.\\

\begin{figure}[h!]
\centering
\includegraphics[width=0.49\textwidth]{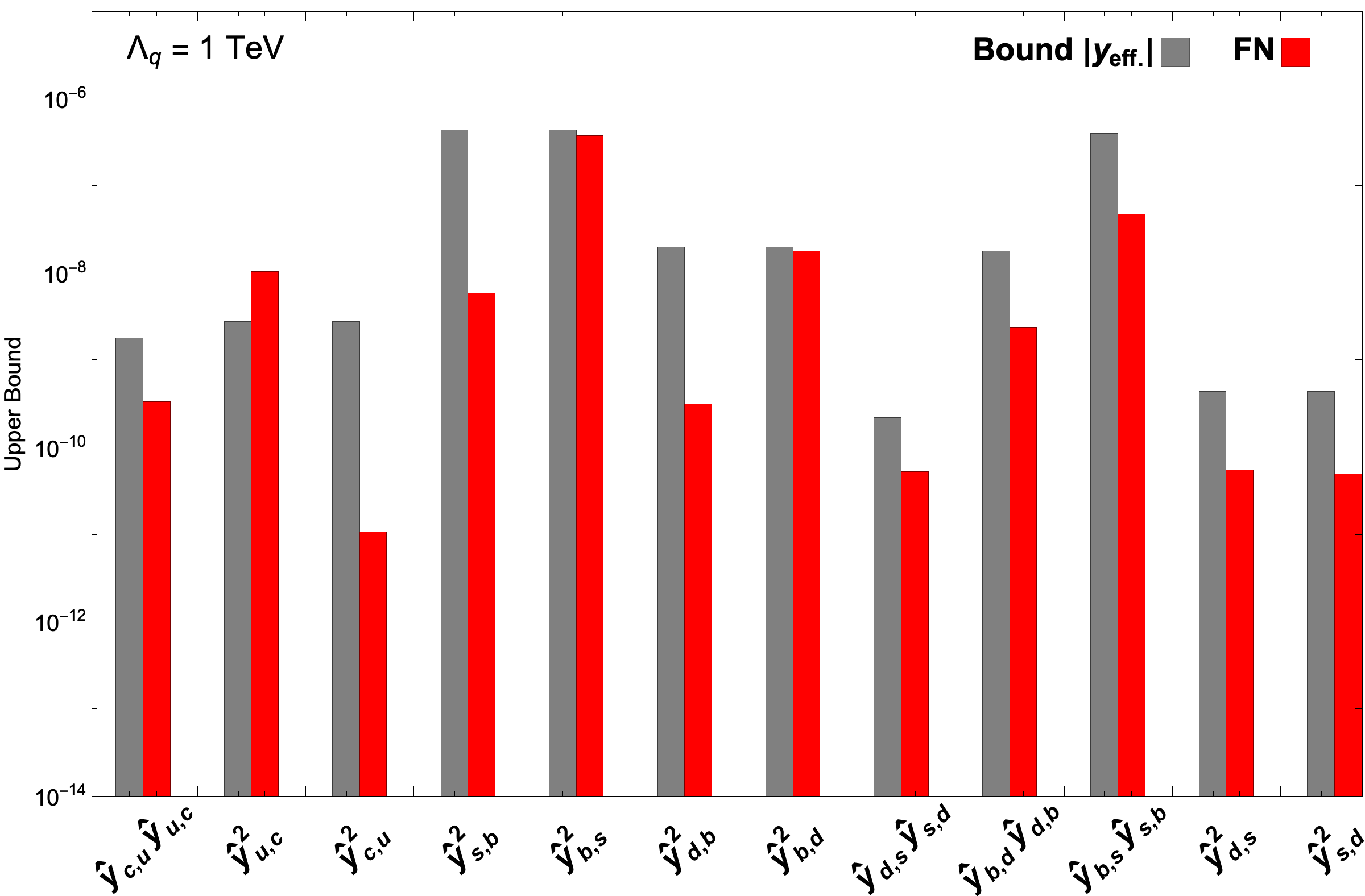}\\
\includegraphics[width=0.49\textwidth]{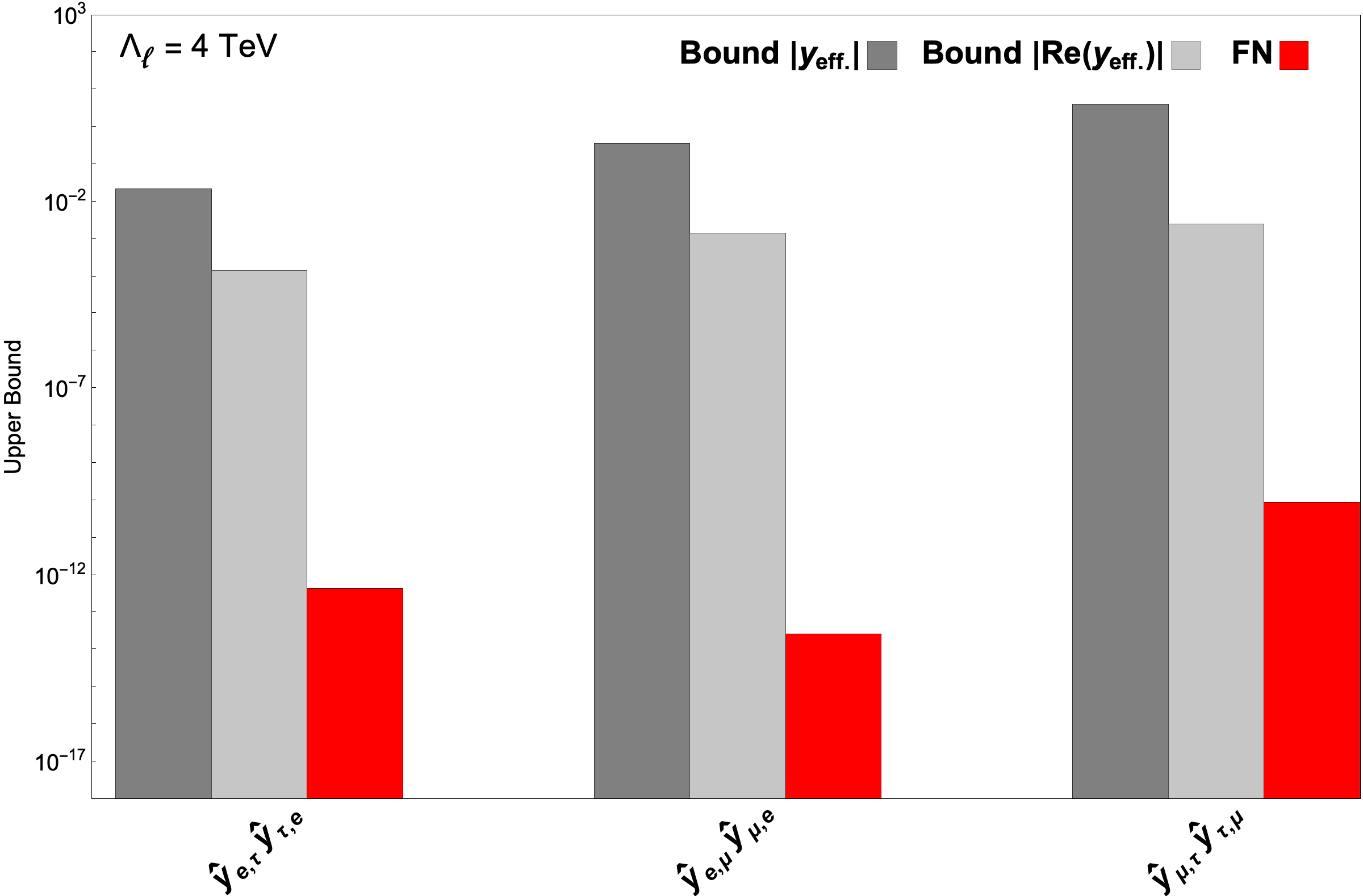}
\includegraphics[width=0.49\textwidth]{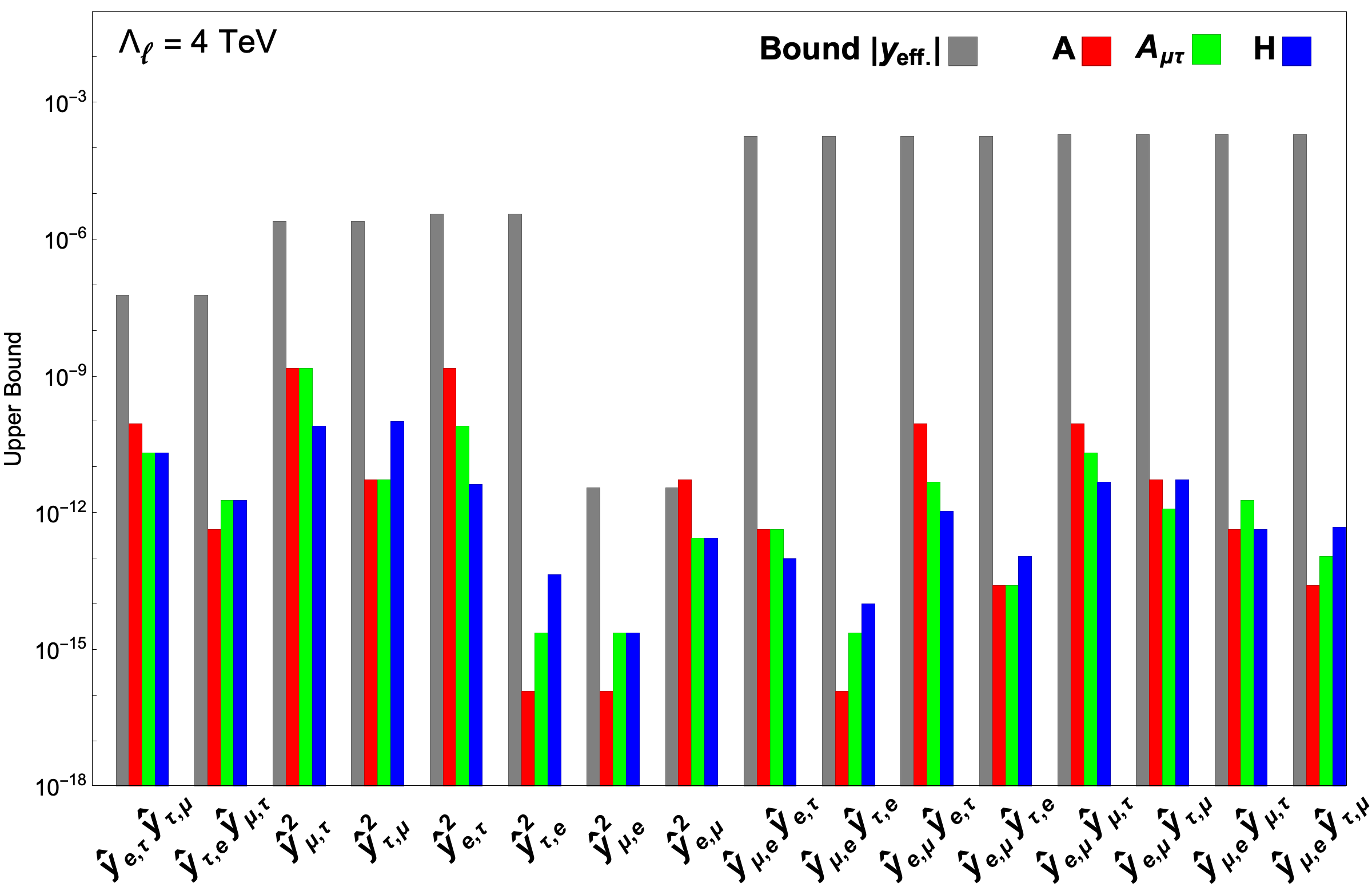}
\caption{\em Graphical representation of the experimental bounds and FN predictions on the effective Yukawas for the CP conserving case. In the first plot the results in the quark sector: in grey the bounds on the absolute value of the combination of the effective Yukawas from Tab.~\ref{tab:QboundsFinal}, and in red the prediction of the FN model assuming $\Lambda_q=1\TeV$. The second plot shows the results in the lepton sector when the combination of the effective Yukawas do not depend on the FN charge assignment: in the dark (light) grey the bounds on the absolute value (real part) from Tab.~\ref{tab:boundsFinal}, while the FN predictions are in red assuming $\Lambda_\ell=4\TeV$. The last plot still refers to the lepton sector and to the effective Yukawas that do depend on the FN charge assignment: in grey there are the bounds on the absolute value from Tab.~\ref{tab:boundsFinal}, in red the Anarchical model, in green the $A_{\mu\tau}$ model and in blue the hierarchical model for $\Lambda_\ell=4\TeV$.}
\label{fig:SummaryOffDiag}
\end{figure}

In Fig.~\ref{fig:SummaryOffDiag}, we give the graphical representation of the experimental bounds and the FN predictions for quarks and leptons with $\Lambda_q=1\TeV$ and $\Lambda_\ell=4\TeV$. These two values are the lower bounds on the scale $\Lambda_q$ and $\Lambda_\ell$ consistent with all the considered experimental constraints when the Wilson coefficients respect the FN symmetries,  for the CP conserving case, that is $\sin\theta=0$ for any effective phase. 

The corresponding bound on the $r_\psi^2$ parameters reads
\be
\begin{aligned}
0.88\lesssim r^2_q&\lesssim 1.12\qquad\text{for any quark}\\
0.99\lesssim r^2_\ell&\lesssim 1.01\qquad\text{for any lepton}\,,
\end{aligned}
\ee
where the lower (upper) bounds are obtained for the $\cos\theta_\psi=-1$ ($\cos\theta_\psi=+1$).  The values corresponding to $\cos\theta_\psi=0$ are $r_q^2=1.004$ and $r_\ell^2=1.00001$

\begin{figure}[h!]
\centering
\includegraphics[width=0.49\textwidth]{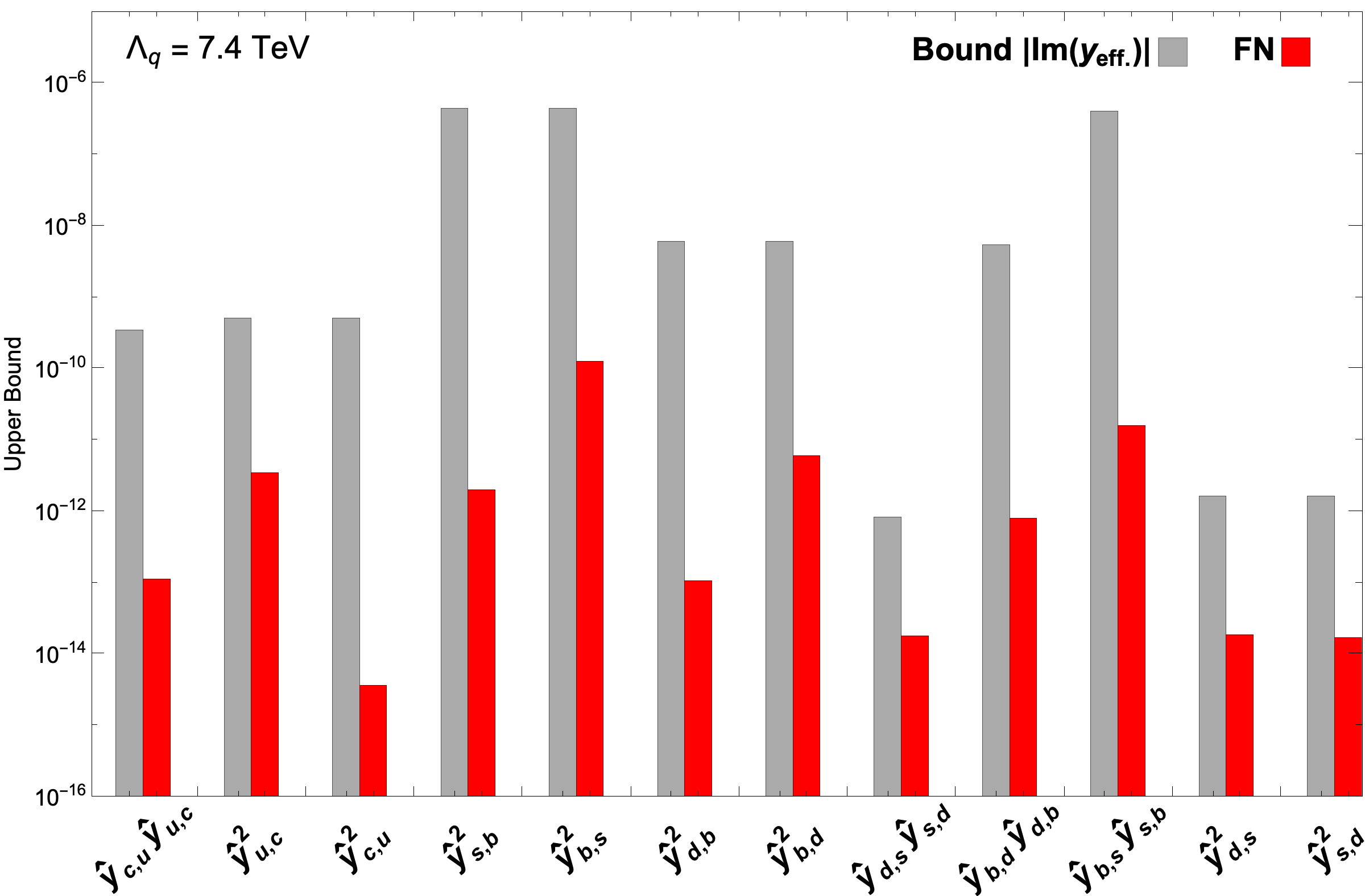}
\includegraphics[width=0.49\textwidth]{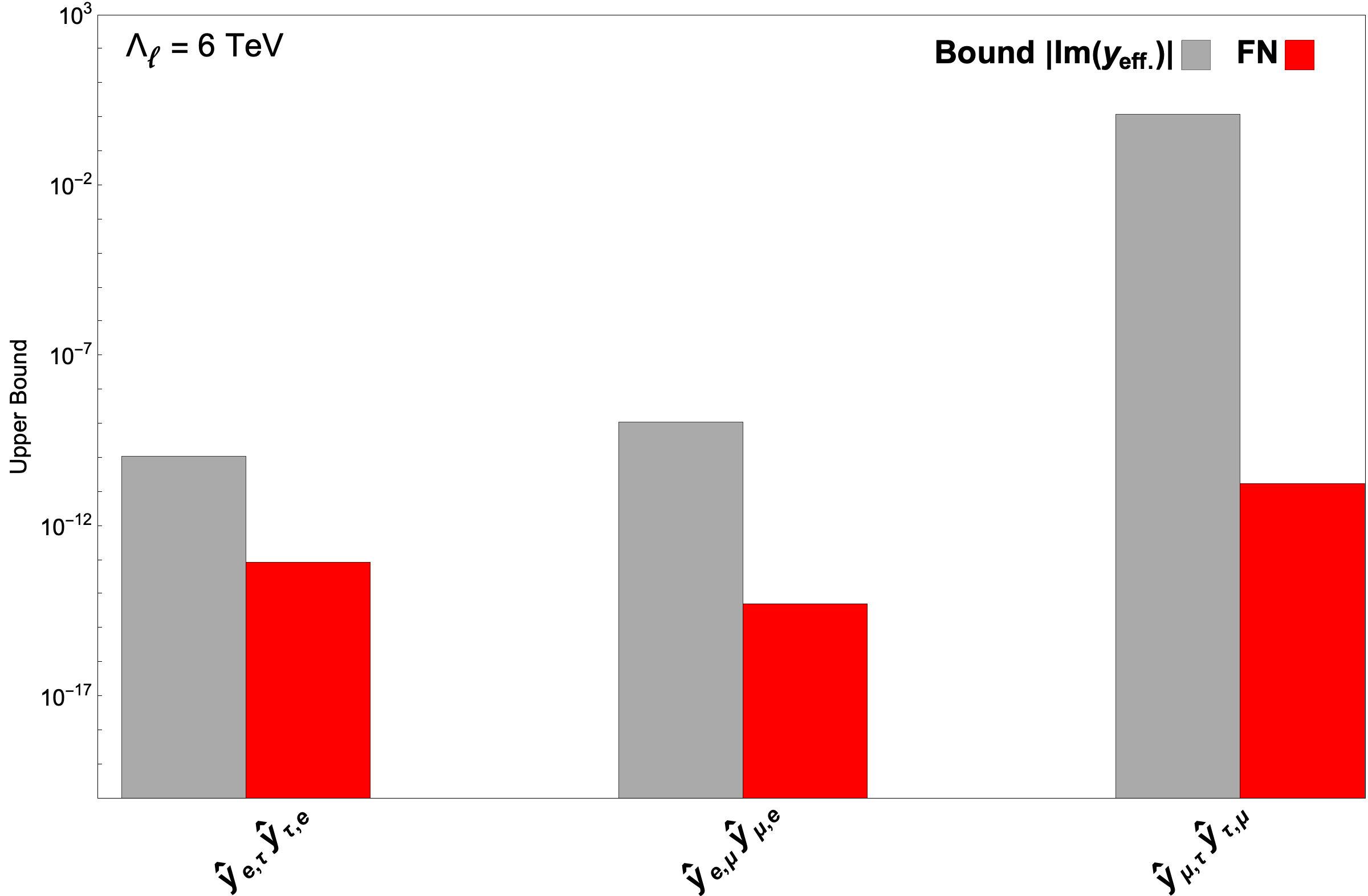}
\caption{\em Graphical representation of the experimental bounds and FN predictions on the effective Yukawas for the case with large CP phases. In the first plot the results in the quark sector: in grey the bounds on the imaginary part of the combination of the effective Yukawas from Tab.~\ref{tab:QboundsFinal}, and in red the prediction of the FN model assuming $\Lambda_q=7.4\TeV$. The second plot shows the results in the lepton sector: in grey the bounds on the Imaginary part from Tab.~\ref{tab:boundsFinal}, while the FN predictions are in red assuming $\Lambda_\ell=6.0\TeV$.}
\label{fig:SummaryOffDiag2}
\end{figure}

The opposite case is the one with large phases of the effective  third generation Yukawas and then the strongest bounds arise from the electron EDM data: in Fig.~\ref{fig:SummaryOffDiag2}, we give the graphical representation of the experimental bounds and the FN predictions for quarks and leptons with $\Lambda_q=7.4\TeV$ and $\Lambda_\ell=6.0\TeV$. As can be seen, once the electron EDMs bounds are satisfied, the FN predictions for the other CPV observables are well below the present experimental limits. 

The corresponding bound on the $r_\psi^2$ parameters reads
\be
\begin{aligned}
0.998\lesssim r^2_q&\lesssim 1.002\qquad\text{for any quark}\\
0.997\lesssim r^2_\ell&\lesssim 1.003\qquad\text{for any lepton}\,,
\end{aligned}
\ee
where the lower (upper) bounds are obtained for the $\cos\theta_\psi=-1$ ($\cos\theta_\psi=+1$).

We can conclude  that if the NP is approximately CP conserving then the $q\ov{q}h$ couplings may still deviate from the SM values by a few percent and the future collider data on the Higgs production and decay  are  very much welcome. The $\ell\ov\ell h$ couplings are much more strongly constrained even in the CP conserving case.  An interesting chance for testing the $\mu \ov{e} h$ couplings is provided by the $\mu\to e $ conversion in nuclei. The present experimental bound is $\BR(\mu\to e N) < 7\times10^{-13}$. Using the scale of $\Lambda_\ell= 4 \TeV$ and FN symmetry, the bound we  get is  $\BR(\mu \to e N) < 3.8 \times10^{-16}$. Future experiments claim to reach a sensitivity of $10^{-16}$~\cite{COMET:2009qeh}, which is  of the same order of magnitude.

In the case of large CP violating phases in the Higgs couplings with $t$, $e$ and $\tau$, all the $h$ couplings are constrained to be very close to their SM values and only higher precision selected flavour data may show the deviations  from the SM, as can be seen in Fig.~\ref{fig:SummaryOffDiag2}.

\section{Flavour Changing Tree-Level Higgs and Top Decays}
\label{sec:TreeLevelDecays}

This section is devoted to the analysis of processes that occur at the tree level due to flavour changing Yukawa couplings: in particular we will focus on flavour changing Higgs decays into leptons or quarks, and on flavour changing top decays. Differently from the observables considered in the previous section, these  decays are safe from NP contributions different from the ones in Eq.~\eqref{LagHMassBasis}. Consistently with the results in the previous section, we will adopt $\Lambda_q=1\TeV$ and $\Lambda_\ell=4\TeV$.

The first observables we consider are flavour violating Higgs decays into fermions, but the top. When the final state particles are leptons, the decay rates are quite strongly constrained experimentally. The explicit expression for the decay width is given by~\cite{Lee:2003nta,Arganda:2004bz}
\begin{equation}
\begin{split}
\Gamma\p{h \to f_i f_j} =& \Gamma\p{h \to \ov{f}_i f_j} +\Gamma\p{h \to f_i \ov{f}_j} =\\
=&\frac{N_C}{16 \pi} m_h\sqrt{\left[1-\left(x_i+x_j\right)^2\right]\left[1-\left(x_i-x_j\right)^2\right]}\times\\
&\times\Bigg[(1 - x_i^2+x_j^2)\p{\absval{\hat{Y}_{f,ij}}^2 + \absval{\hat{Y}_{f,ji}}^2}- 4\, x_ix_j\, \Re\left(\hat{Y}_{f,ij} \hat{Y}_{f,ji}\right)\Bigg] \, ,
\end{split}
\label{eq:Higgs_FV_width}
\end{equation}
for $i\neq j$, $N_C=3$ for quarks and $N_C=1$ for leptons, $x_i\equiv m_{f_i}/m_h$, and where $\hat{Y}_f$ are defined in Eq.~\eqref{hatYf}. From Eq.~\eqref{ExplicitCFN}, we get
\begin{equation}
\begin{split}
& \absval{C_{f,ij}}^2 \simeq \OO(1) \, Y^2_{f,j}\,\epsilon^{2(n_{Q_i}-n_{Q_j})} \qquad\text{for $f=u,d$}\, , \\
& \absval{C_{e,ij}}^2 \simeq \OO(1) \, Y^2_{e,j}\,\epsilon^{2(n_{L_i}-n_{L_j})}\,, \\
& \Re\left[C_{f,ij} C_{f,ji}\right]\simeq \OO(1)\, Y_{f,i} Y_{f,j} \, \cos \p{\theta_{f,ij}+\theta_{f,ji}} \,,
\end{split}
\label{eq:FN_C_elements_in_FV_Width}
\end{equation}
and an explicit computation of the two terms in the last line of Eq.~\eqref{eq:Higgs_FV_width} reveals that the first term is always dominating for any set $\{f_i,f_j\}$. Neglecting the fermion masses with respect to the Higgs one, Eq.~\eqref{eq:Higgs_FV_width} reduces to
\begin{align}
\Gamma\p{h \to f_i f_j} &\simeq \frac{3}{8 \, \pi} m_h\, \dfrac{v^2}{\Lambda^4}
\left[\OO(1) \,M^2_{f,j}\, \epsilon^{2(n_{Q_i}-n_{Q_j})} + \OO(1) \, M^2_{f,i}\,\epsilon^{-2(n_{Q_i}-n_{Q_j})}\right]\quad \text{for $f=u,d$}\,,\nn\\
\Gamma\p{h \to e_i e_j} &\simeq \frac{1}{8 \, \pi} m_h\, \dfrac{v^2}{\Lambda^4}
\left[\OO(1) \,M^2_{e,j}\, \epsilon^{2(n_{L_i}-n_{L_j})} + \OO(1) \, M^2_{e,i}\,\epsilon^{-2(n_{L_i}-n_{L_j})}\right] \,.
\label{eq:Higgs_FV_width_FN}
\end{align}
The branching ratios for these decays can be obtained dividing this last expression by the Higgs total decay width deduced from Eq.~\eqref{eq:total_width_all}. The results can be read in Tab.~\ref{tab:boundsHtoee} for decays in both quarks and leptons, taking all the $\OO(1)$ parameters equal to $1$. As we can see, the predicted values for all the decays are much below the present experimental limit. 

Notice that for the decays into quarks, the proper observables involve jets or mesons in the final state\cite{Kagan:2014ila}. It is however beyond the scope of this paper to enter into further details. 

\begin{table}[h!]
\centering{
\renewcommand{\arraystretch}{1.7}
\begin{tabular}{|c||c|c|c|c|}
\hline
{\bf BR} & {\bf Experimental Bound \@95\%C.L.} & \multicolumn{3}{|c|}{\bf FN Prediction} \\ 
\hline
\hline
$h\to uc$ & $-$ & \multicolumn{3}{|c|}{$6\times10^{-8}$} \\
$h\to ds$ & $-$ & \multicolumn{3}{|c|}{$6\times10^{-10}$} \\
$h\to db$ & $-$ & \multicolumn{3}{|c|}{$4\times10^{-8}$} \\
$h\to sb$ & $-$ & \multicolumn{3}{|c|}{$8\times10^{-7}$} \\
\hline
\hline
\hline
{\bf BR} & {\bf Experimental Bound \@95\%C.L.} & $A$ & $A_{\mu\tau}$ & $H$ \\ 
\hline
\hline
$h\to e\mu$ & $6.1\times 10^{-5}\text{~\cite{ATLAS:2019old}}$ & $3\times10^{-9}$ & $10^{-10}$ & $1\times10^{-10}$ \\
$h\to e\tau$ & $2.2\times 10^{-3}\text{~\cite{CMS:2021rsq}}$ & $8\times10^{-7}$ & $4\times10^{-8}$ & $2\times10^{-9}$ \\
$h\to \mu\tau$ & $1.5\times 10^{-3}\text{~\cite{CMS:2021rsq}}$ & $8\times10^{-7}$ & $8\times10^{-7}$ & $9\times 10^{-8}$ \\
\hline
\end{tabular}
\caption{\em Branching ratios for the flavour changing Higgs decays. In the second column we report the experimental upper bounds when available, while in the other columns the FN predictions, specifying the leptonic FN realisation. The cut-off scales are fixed to $\Lambda_q=1\TeV$ and $\Lambda_\ell=4\TeV$ and all the $\OO(1)$ parameters are taken equal to $1$.}
\label{tab:boundsHtoee}}
\end{table}

The last observable we will consider in this section is flavour changing top decays, $t\to hc$ and $t\to h u$. The explicit expression for the decay width is given by~\cite{Hou:1991un}
\be
\Gamma\p{t\to h u_i}=\frac{1}{32\pi}m_h\left|\hat{Y}_{u,i3}\right|^2\left[\left(1+x_i\right)^2-x_h^2\right]\sqrt{1-(x_h+x_i)^2}\sqrt{1-(x_h-x_i)^2}\,,
\ee
where $x_i\equiv m_{u_i}/m_t$ and $x_h\equiv m_h/m_t$. Considering the definition of $\hat{Y}_u$ and the expression for $C_{u,ij}$ in Eq.~\eqref{eq:FN_C_elements_in_FV_Width}, we get
\be
\left|\hat{Y}_{u,i3}\right|^2\simeq\OO(1)\dfrac{2m_t^2v^2}{\Lambda_q^4}\epsilon^{2(n_{Q_i}-n_{Q_3})}
\ee
and the generic expression, neglecting $x_i$ with respect to $x_h$, reduces to
\be
\Gamma\p{t\to h u_i}\simeq\frac{\OO(1)}{16\pi}m_h\dfrac{m_t^2v^2}{\Lambda_q^4}\epsilon^{2(n_{Q_i}-n_{Q_3})}\left(1-x_h^2\right)^2\,.
\ee
The numerical results for the corresponding branching ratios, considering the total decay width $\Gamma_t=1.42^{+19}_{-15}\GeV$~\cite{Zyla:2020zbs}, can be read in Tab.~\ref{tab:boundsttohuc}: the FN predictions are below the present experimental limits of at least two orders of magnitude.

\begin{table}[h!]
\centering{
\renewcommand{\arraystretch}{1.7}
\begin{tabular}{|c||c|c|}
\hline
{\bf BR} & {\bf Experimental Bound \@95\%C.L.} & {\bf FN Prediction} \\ 
\hline
\hline
$t\to hu$ 					& $4.5\times 10^{-3}\text{~\cite{ATLAS:2015ncl}}$ 	& $2\times10^{-6}$  \\
$t\to hc$ 	& $4.6\times 10^{-3}\text{~\cite{ATLAS:2015ncl}}$, $5.6\times 10^{-3}\text{~\cite{CMS:2014qxa}}$	& $4\times10^{-5}$ \\
\hline
\end{tabular}
\caption{\em Branching ratios for the flavour changing top decays. In the second column we report the experimental upper bounds, while in the third column the FN predictions. The cut-off scale is fixed to $\Lambda_q=1\TeV$ and all the $\OO(1)$ parameters are taken equal to $1$.}
\label{tab:boundsttohuc}}
\end{table}

\section{Conclusions}
\label{sec:concl}

In contrast to the electroweak part of the Standard Model, beautifully understood in terms of the underlying gauge symmetries, its flavour sector is very mysterious. It relies on purely phenomenological description  encoded in the fermion masses and mixings, which originate from the Yukawa coupling  matrices introduced at the level of the $SU(2)_L\times U(1)_Y$ invariant renormalisable Lagrangian, with the SM fermion spectrum. The latter are however beyond any SM theoretical control. Better theoretical understanding of the flavour physics is one of the main  BSM challenges, parallel to the naturalness problem. It  may well be that the scales of BSM physics that can  shed some light on the flavour problem are very different from the BSM physics linked to the Brout-Englert-Higgs mechanism.

In this paper we have determined the lower bounds on the scale of new physics possibly contributing to the $f\ov{f}h$ effective couplings following from the variety of processes dependent  on those couplings. We have worked in the framework of SMEFT and our main assumption is that the Wilson coefficients of the relevant dim 6 operators respect certain flavour structure:  either the Minimal Flavour Violation ansatz or  a flavour symmetry, often invoked  to explain  the observed pattern of fermion masses and mixings. Thus, our results determine the sensitivity of different observables to the flavour BSM physics under the mentioned above assumption. 

There are several sources of experimental information on the effective Yukawa couplings. The absolute values of the third generation couplings (and some of the second one) are  directly measured in the Higgs boson production and decays at the LHC experiments.  Their imaginary parts are bounded  by the limits on EDMs. The non-diagonal couplings are bounded by a vast number of FCNC processes and radiative decays. The latter bounds are conservative as those processes may be subject also to other new physics contributions at the same level of precision calculations but we take them on equal footing with the collider bounds.

The emerging picture of the bounds on new physics scales is discussed in detail in Sect.~\ref{sec:OffDiag}, while the predictions for flavour changing Higgs and top decays in Sect.~\ref{sec:TreeLevelDecays}. Here are the main points:
\begin{itemize}
\item[-]
First,  we disregard the bounds on the imaginary parts of the Yukawas (assuming that they are controlled by approximately CP conserving BSM theory). With the present precision, the FCNC data give  stronger bounds on the scale of NP than the collider data (obviously this does not apply to the MFV ansatz, where the off-diagonal couplings are absent and only collider data are relevant). We have then $\Lambda_q = 1(0.8)$ TeV and $\Lambda_\ell = 4(0.5)$ TeV for FCNC(collider) data.
\item[-] 
The low FCNC bound  for quarks  is consistent with a few percent deviations from the SM in the Yukawa couplings and  high precision collider data may show such deviations. For leptons, the FCNC bound is stronger. Its violation by the collider data would mean the violation of the flavour symmetry or the presence of additional NP affecting  the flavour observables that leads to cancellations. 
\item[-]
Flavour changing Higgs and top decays are predicted to be at least two orders of magnitude below the present experimental limits. In this case, a violation of those bounds would univocally mean the violation of the flavour symmetry.
\item[-]
A promising observable is the $\mu\to e $ conversion in nuclei. Future experiments claim to reach a sensitivity of $10^{-16}$~\cite{COMET:2009qeh}, which is  of the same order of magnitude as our predicted upper bound.
\item[-]
In the extreme case, if the phases in $\tilde{\kappa}_{t,e,\tau}$ are maximal, then $\Lambda_q= 7.4\TeV$  and $\Lambda_\ell = 6\TeV$ from the upper limit on the electron EDM and only selected CP violating observables may show
some deviations from the SM predictions, provided the precision  of the data improve by an order of magnitude (see Fig.~\ref{fig:SummaryOffDiag2}).
\end{itemize}

\section*{Acknowledgements}

L.M. and A.d.G. thanks Emmanuel Stamou for discussions. J.A.G., A.d.G. and L.M. acknowledge partial financial support by the Spanish MINECO through the Centro de excelencia Severo Ochoa Program under grant SEV-2016-0597, by the Spanish ``Agencia Estatal de Investigac\'ion''(AEI) and the EU ``Fondo Europeo de Desarrollo Regional'' (FEDER) through the project PID2019-108892RB-I00/AEI/ 10.13039/ 501100011033, by the European Union's Horizon 2020 research and innovation programme under the Marie Sk\l odowska-Curie grant agreement No 860881-HIDDeN. L.M. acknowledges partial financial support by the Spanish MINECO through the ``Ram\'on y Cajal'' programme (RYC-2015-17173). A.d.G. acknowledges financial support by the ETH Z\"urich and the Institut f\"ur Theoretische Physik. The research of S.P. has received funding from the Norwegian Financial Mechanism for years 2014-2021, grant nr 2019/34/H/ST2/00707. S.P thanks Anna Lipniacka (University of Bergen)  for discussions on the Higgs boson LHC data.


\footnotesize


\providecommand{\href}[2]{#2}\begingroup\raggedright\endgroup

\end{document}